\documentclass{aa}

\usepackage{graphicx}
\usepackage{natbib}
\usepackage[dvipsnames]{xcolor}
\usepackage{booktabs}
\usepackage{txfonts}

\newcommand{\lama}{$\lambda$\,And}
\newcommand{\cahk}{Ca\,{\sc ii}\,H\&K}
\newcommand{\cairt}{Ca\,{\sc ii}\,IRT}
\newcommand{\Halpha}{H$\alpha$}
\newcommand{\Hbeta}{H$\beta$}
\newcommand{\kms}{km\,s$^{-1}$}
\newcommand{\ms}{m\,s$^{-1}$}
\newcommand{\ergs}{erg\,cm$^{-2}$s$^{-1}$}

\begin{document}

\title{First Doppler image and starspot-corrected orbit for $\lambda$~Andromedae} 
\subtitle{A multifaceted activity analysis}

\author{\"O. Adebali\inst{1,2}, K. G. Strassmeier\inst{1,2},
I. V. Ilyin\inst{1}, M. Weber\inst{1}, D. Gruner\inst{1} \and Zs. K\H{o}v\'ari\inst{3,4}}

\institute{
Leibniz-Institute for Astrophysics Potsdam (AIP), An der Sternwarte 16, D-14482 Potsdam, Germany\\
\email{oadebali@aip.de}
\and
Institut f\"ur Physik und Astronomie, Universit\"at Potsdam, D-14476 Potsdam, Germany
\and
HUN-REN Research Centre for Astronomy and Earth Sciences, Konkoly Observatory, H-1121 Budapest, Hungary
\and
HUN-REN CSFK, MTA Centre of Excellence, H-1121 Budapest, Hungary
                }

   \date{Received 19 November 2024 / Accepted 08 February 2025}

  \abstract
   {Starspots on a rotating stellar surface impact the measured radial velocities and thereby limit the determination of precise orbital elements as well as astrophysical stellar parameters and even jeopardize the detection and characterization of (exo)planets.}
   {We quantify the effect of starspots for the orbital elements of the spotted RS\,CVn binary \lama\ and present an empirical correction. The aim is to obtain a more precise orbital solution that can be used to better study the system's severe orbital-rotational asynchronism.}
   {Phase-resolved high-resolution optical spectra were recorded over the course of 522 days in 2021-2022. We employed two facilities with medium and high resolution spectroscopy for the multiple activity analyses. Doppler imaging is used to reconstruct \lama's starspots with a high resolution ($R=$\,250\,000) and high signal-to-noise ratio spectra. Optimized cross-correlation functions were used to measure precise radial velocities at a level of a few ten's of m/s. }
   {The spot-corrected radial velocities enable, on average, a threefold increase in precision of the individual orbital elements. The residual velocity jitter with a full range of 500\,\ms\ is modulated by the rotation period of \lama\ of 54.4$\pm$0.3\,d. Our logarithmic gravity from spectrum synthesis of 2.8$\pm$0.2 together with the interferometrically determined stellar radius suggest a most-likely mass of the primary of $\approx$1.4\,M$_\sun$. The small orbital mass function then implies a secondary mass of just $\approx$0.1\,M$_\sun$, which is appropriate for an L-class brown dwarf. The Doppler image reconstructs a dominating cool spot with an umbral temperature difference of $\approx$1000\,K with respect to the photosphere of 4660\,K and is likely surrounded by a moat-like velocity field. Three more weaker spots add to the total surface spottedness, which is up to 25\% of the visible surface. Seven optical chromospheric tracers show rotational modulation of their emission line fluxes in phase with the cool spots. This surface configuration appears to have been stable for the 522 days of our observations. We also redetermined the carbon isotope ratio to $^{12}\text{C}/^{13}\text{C}=30\pm5$ and measured a contemporaneous disk-integrated mean longitudinal magnetic field of polarity Plus/Minus up to $2.70\pm0.35$\,G from Stokes-V line profiles.}
   {}


\authorrunning{Adebali et al.}


   \maketitle
%

\section{Introduction}

The RS~Canum~Venaticorum (RS\,CVn) binaries are cool, evolved, close binary systems with an exceptionally high level of magnetic activity. These systems were originally defined by \citet{Hall1972} and further specified in \citet{Hall1976}, \citet{Vogt1983}, and \citet{Strassmeier1993}. Many of these RS~CVns were cornerstones for studying the relationship between the global magnetic activity and the photospheric starspots \citep[e.g.,][]{Strassmeier2009}. \lama\ is such a cornerstone RS\,CVn system and it is very attractive to observers because of its brightness of $V$=3.82~mag \citep{Ducati2002}. It is a single-lined spectroscopic binary (SB1) with a subgiant of classification G8III-IV and a surface temperature of $\approx$4600\,K \citep{Savanov&Berdyugina1994}. It is approximately of solar mass with a radius of around 7\,$R_\odot$ \citep{Drake2011}. The rotation period of 54\,d of the primary star was determined from photometry and is among the longest known for RS\,CVn binaries \citep{Landis1978, Strassmeier1989, Henry1995}. \lama's rotational line broadening ($v\sin i$) is thus comparably small, approximately 7\,\kms. But it is the orbital period of $\approx$20\,d and the nearly circular orbit that makes \lama\ outstanding because the orbital revolution and the (primary-star) rotation are grossly asynchronous. In this respect, it is even more surprising that the current best orbit for \lama\ dates back to the 1940s. \citep{Walker1944}.

\citet{Mirtorabi2003}'s study of the long-term (1976–2002) photometric behavior of \lama\ showed evidence of a complex starspot-faculae activity cycle,  with timescales as
short as 4~yr and as long as 14~yr. At least during the 5~yr
interval in which they presented TiO photometry, it
appears that most of the long-term changes in the star’s
brightness were produced by variations in the
fractional areal coverage of the bright regions.

To quantify \lama's spatial surface activity, photometric \citep{Donati1995}, spectropolarimetric \citep{Fionnagain2021}, and interferometric \citep{Parks2021,Martinez2021} imaging techniques have been applied. While \citet{Donati1995} published the first surface map of the star from $BV$ light curve inversions, \citet{Parks2021} and \citet{Martinez2021} obtained direct surface images for two epochs from H-band interferometry in 2010 and 2011. These images showed several spots at near-equatorial latitudes and predominantly only in one hemisphere. \citet{Fionnagain2021} applied the Zeeman-Doppler imaging (ZDI) technique to Stokes-V line profiles and showed that the star possesses an equatorial magnetic field with radial and azimuthal components of local strengths of up to 83\,G. Because the ZDI technique relies on the differential circularly polarized line profiles rather than the integral-light profiles (i.e. Stokes-I), it is applicable even to stars with a comparably small $v\sin i$, such as \lama. Doppler imaging (DI), on the other hand, has not yet been attempted for \lama\ because its low $v\sin i$ requires very high spectral resolution in order to sample the narrow line profiles properly (such that a wavelength point in the line profile can be related to a surface pixel via the Doppler effect). Nevertheless, if applicable, DI is still the most powerful method to indirectly infer spatial information on a stellar surface. It not only allows the number, size, temperature, and morphology of starspots to be reconstructed, but also their growth and decay and their migration in latitude and longitude. Furthermore, determining the surface morphology gives us a chance to understand the reasons for the activity effects on the other properties of the star, such as radial velocity (RV) jitter, surface temperature evolution, and chromospheric emission.

The effects of starspots on stellar RVs have been investigated by many authors in the recent literature, in particular in connection with the search for exoplanets. Early simulations by \citet{Saar&Donahue1997} showed that starspots can create RV jitter up to 200\,\ms. Similar to this work, \citet{Hatzes2002} modeled the RV perturbations as a function of spot filling factors and for a range of rotational velocities up to $v\sin i$ of 13\,\kms. \citet{Hatzes2002} showed that the amplitude of the RV jitter is directly proportional to both the $v\sin i$ and the spot filling factor. \citet{Desort2007} simulated RV curves and line bisectors by placing starspots at different latitudes of the stellar disk of a solar-like star with different inclinations of the rotational axis. However, all of these simulations considered only a single spot on the surface. \citet{Boisse2011} used a more realistic two-spot configuration with the spots at different latitudes and longitudes. They demonstrated similarities and discrepancies based on a periodogram analysis of the residual RV perturbations. \citet{Rajpaul2015} obtained RV residuals by applying Gaussian processes that included an exoplanet signal in the model. An observational study by \citet{Donati2016} demonstrated the effects of stellar surface structures on the RV curve of a T-Tauri star. By filtering out the activity, they discovered the residual exoplanet signal. Similar results had been achieved for the secondary component of the pulsation-dominated RV data of the M-supergiant Betelgeuse \citep{MacLeod2025}. Also recently, \citet{Zhao2023} investigated the effects of (bright) faculae and its combined effect with (dark) starspots.  

In the current paper, we present our new observations and the data processing (Sect.~\ref{Observation_Reduction}) together with an empirical attempt for separating the spot-related RV from the orbit-related RV and thereby refining \lama's orbital elements to a higher precision (Sect.~\ref{Data_Analysis}). In addition, we introduce the first Doppler image of \lama\ and compare it with the RV perturbations and the results from other mapping attempts in Sect.~\ref{DI_section}. Moreover, in Sect.~\ref{ca}, we analyze the star's  chromospheric activity and relate it to the photospheric properties of the star. A discussion and summary in Sect.~\ref{Discussion} is followed by our conclusions in Sect.~\ref{Conclusion}.

\begin{figure*}
    \centering
    \includegraphics[width=\linewidth]{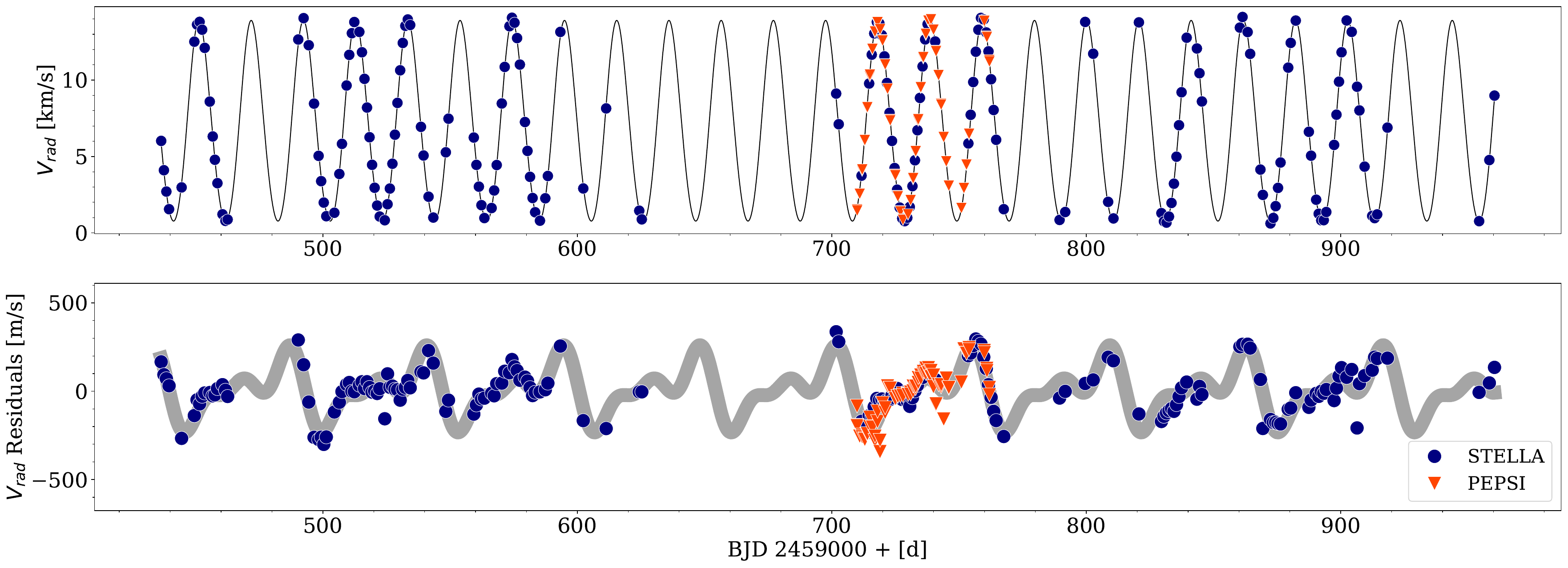}
    \caption{Radial velocities of $\lambda$\,And from STELLA+SES and VATT+PEPSI. In the upper panel, the orbital fit is plotted as a black line along with the data. STELLA+SES observations are indicated with dark blue circles; VATT+PEPSI observations are plotted in red triangles. In the lower panel the RV residuals are shown after removing the predicted orbital velocities. The thick grey line emphasizes the rotational modulation by applying  a three-sinusoidal fit to the RV residuals.  }
    \label{rvs_res}
\end{figure*}

\section{Observations and data reduction}\label{Observation_Reduction}

We collected high-resolution, high signal-to-noise (S/N) spectroscopic data with the 1.2-m STELLA (STELLar Activity) robotic telescope \citep{stella2004} in Tenerife, Spain, equipped with the STELLA Echelle Spectrograph (SES), and with the 1.8-m Vatican Advanced Technology Telescope (VATT) in southern Arizona, United States, fiber-fed with the Potsdam Echelle Polarimetric and Spectroscopic Instrument \citep[PEPSI;][]{Strassmeier2015} from the Large Binocular Telescope (LBT). STELLA data were gathered autonomously once per (clear) night between January 2021 and July 2022, while the VATT was employed manually for a dedicated run between UT~May~10, 2022 and July~1, 2022, covering a full stellar rotation of \lama. A total of 201 STELLA spectra and 81 VATT spectra were acquired. On UT~Oct.~22, 2024, Oct.~15, 2024, Jan.~04, 2022 and Oct.~13, 2017 we obtained deep Stokes-IQUV spectra of \lama\ for 3837--9067\,\AA\ with the PEPSI polarimeter on the LBT. The circular polarization spectra were obtained with the use of a super-achromatic quarter-wave retarder and the Foster prism as a polarizing beam-splitter. The spectra were observed at two angles separated by 90 degrees and reduced with the difference method \citep{ilyin2012} which resulted in the Stokes I and V spectra discussed in Sect 5.3. The polarimetric spectra have a 2-pixel resolution of 130\,000 and a cumulative S/N from all six subexposures of around 5\,000 in Stokes-I in the red wavelengths. 

SES is a fiber-fed fixed-format echelle spectrograph providing a 3-pixel resolution of 55\,000 over a wavelength range of 3900-8800\,\AA. At a wavelength of 6000\,\AA\ this corresponds to an effective resolution of 5.5\,\kms\ or 110\,m\AA . Integration time was 280\,s and resulted in an average S/N of 400 per pixel. Its detector is an e2v 4k$\times$4k CCD with 15\,$\mu$m pixels. The spectral resolution is made possible with a 2-slice image slicer and a 67-$\mu$m (octagonal) fiber with a projected sky aperture of 3.8 arc seconds. The spectrograph is placed in a thermal housing but not in a pressure sealed environment. Its RV stability is thus on average 50\,\ms\ \citep{vpnep}. SES spectra are automatically reduced using the IRAF-based SES data-reduction (SESDR) pipeline \citep[see][]{weber16}. The observing log is given in Table~\ref{Table_STELLA-SES}.

PEPSI is a fiber-fed stabilized echelle spectrograph with two arms (blue and red) and three cross dispersers (CD) per arm. The spectrograph is located in the basement of the LBT with an underground fiber connection from the VATT of length 450\,m. We used it in its high-resolution 2-pixel $R$=250\,000 mode (called 100L mode) with two pairs of CDs (CD3+CD5 and CD3+CD6) covering the wavelength ranges 4800-5400\,\AA\ and 6278-9067\,\AA. Its detectors are two STA 10k$\times$10k CCDs with 9\,$\mu$m pixels. The high resolution is achieved with a pair of 100\,$\mu$m fibers and a 9-slice image slicer. The spectrograph is kept stable in a thermal and pressure housing enabling a long-term RV precision of 15\,\ms\ when fed through the VATT fiber connection \citep{vpnep}. PEPSI spectra were reduced with the SDS4PEPSI pipeline based on \citet{ilyin4A}; see also \citet[][]{Str2018}. Integration time was 40\,min and achieved S/N of on average 400 in CD3, and 900 in CD5 and CD6 (per pixel). The observing log is given in Table~\ref{Table_PEPSI-VATT}.

\section{Radial velocity analysis} \label{Data_Analysis}

\subsection{Toward a RV measurement}

Radial velocities are obtained from a fit of the cross-correlation function (CCF) of the observed spectra with a synthetic template. We followed the same procedure recently described for the VPNEP survey \citep{vpnep} and we refer the reader to this paper for a detailed description. For \lama, we use a Phoenix spectrum as the only template with $T_{\rm eff}$/$\log g$/[Fe/H] of 4600/3/-0.5 from \citet{Husser2013} broadened to match the full line width of \lama\ of $\approx$8.7\,\kms. While SES spectra cover the entire optical wavelength range, the sections for the CCF were nevertheless restricted by excluding  regions of strong lines and wavelengths longer than the terrestrial O$_2$ A-band at 7600\,\AA. For PEPSI, only the CD-3 wavelength range (4800-5400\,\AA) was used for the CCF because it contains the majority of lines and is basically free of terrestrial features. We see no trace of a secondary star in the CCFs of primary line profiles.

Because STELLA is a robot and tries to observe at any allowed observing condition, it also produces low-quality spectra during nights of bad seeing and/or clouds. We thus applied a 3$\sigma$ clipping for removing the outliers in our RV data. A total of seventeen RVs were thereby removed, all of them from very low S/N spectra.

During the VPNEP survey, \citet{vpnep} used stars that were commonly observed with SES and PEPSI to determine their zero-point offset between the two telescope-spectrograph combinations. The grand mean RV difference for SES-minus-PEPSI and its standard deviation was $-$395$\pm$209\,\ms. As seen, it had a rather large RMS mostly due to the large range of spectral types of the joint stars. For \lama\ the difference is consistently just $-$180\,\ms, which we removed from the SES RVs. All RVs in this paper are reduced to the barycentric motion of the solar system using the JPL ephemeris. Tables~\ref{Table_STELLA-SES} and \ref{Table_PEPSI-VATT} in the appendix list the RVs for all spectra.

\begin{figure}
    \centering
    \includegraphics[width=9 cm]{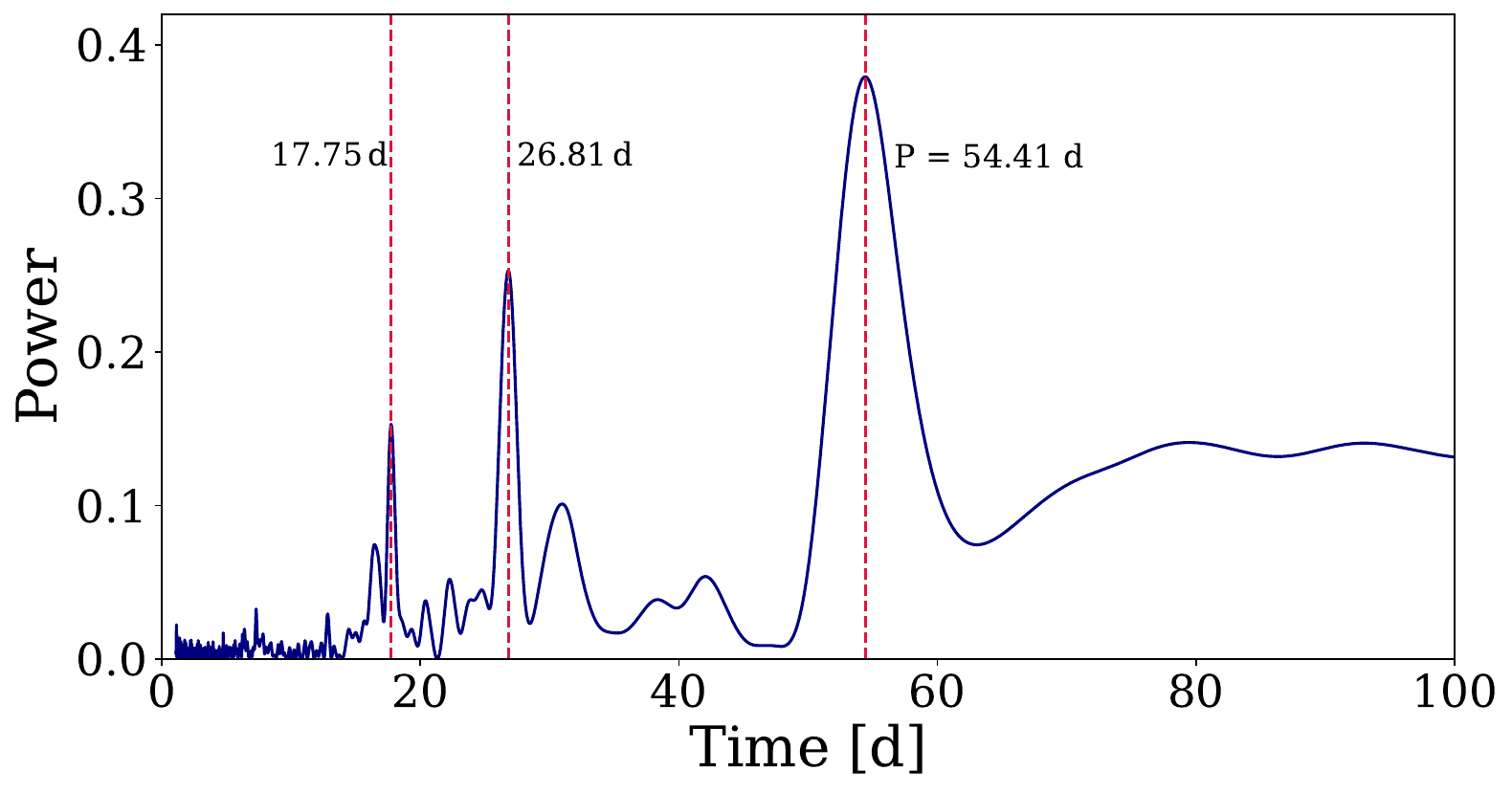}
    \caption{Lomb-Scargle periodogram for the RV-residuals from Fig.~\ref{rvs_res}. The strongest peak at 54.41\,d is attributed to the rotation period of the primary component. Two other peaks are found near $P/2$ and $P/3$.}
    \label{LS1}
\end{figure}

\subsection{Periods and phases}

We apply both the Lomb Scargle Periodogram \citep[LSP;][]{Lomb1976,Scargle1982} and the Phase Dispersion Minimization \citep[PDM;][]{pdm} for the determination of the change in the activity indices properly. The LSP as a variant of a Fourier transformation is the least-squares fitting of a harmonic (sinusoidal) function to a time series. As shown in Fig.~\ref{rvs_res} the RV series is nearly sinusoidal and well sampled, and thus well suited for the Lomb Scargle approach. We see period errors of well below 1\%\ from the covariance matrix of the RV fit of a single harmonic with a particular period. 

For the periodogram of the RV residuals (after removing the predicted velocity from the orbit) we apply a Monte-Carlo approach by selecting randomly 64\%\ of the RV data points (170 data points) and, from this subsample, again randomly selecting 95 RV data points (which is the difference 36\%) which then together makes up an equally sized sample compared to the original. This is repeated 10,000 times and a LSP computed for each. Its RMS is adopted as the most-likely error. The PDM method is similar to a $\chi^2$ reduction when fitting data models. Frequencies were searched at an equidistant sampling between 0.0001 and 1\,c/d.

Because of the strong asynchronism between orbital and rotational motion of \lama\ its observational phases are determined with two different periods. In this paper, orbital epochs and phases, $E.\phi$, are computed from times of observation (in Barycentric Julian Date \textit{BJD}) via
\begin{equation} \label{eq1}
BJD = T_0 + P_{\rm orb} \times\ E.\phi \ ,
\end{equation}
while rotational epochs and phases, $E.\varphi$, are computed from
\begin{equation} \label{eq2}
BJD = T_0 + P_{\rm rot} \times\ E.\varphi \ .
\end{equation}
Throughout the paper, we use $P_{\rm orb}$=20.520394$\pm$0.000041\,d and a time of periastron passage $T_0$=2,459,696.372 from our new orbit, and $P_{\rm rot}$=54.41$\pm$0.3\,d from its RV residuals.

\begin{figure}
    \includegraphics[width=\linewidth]{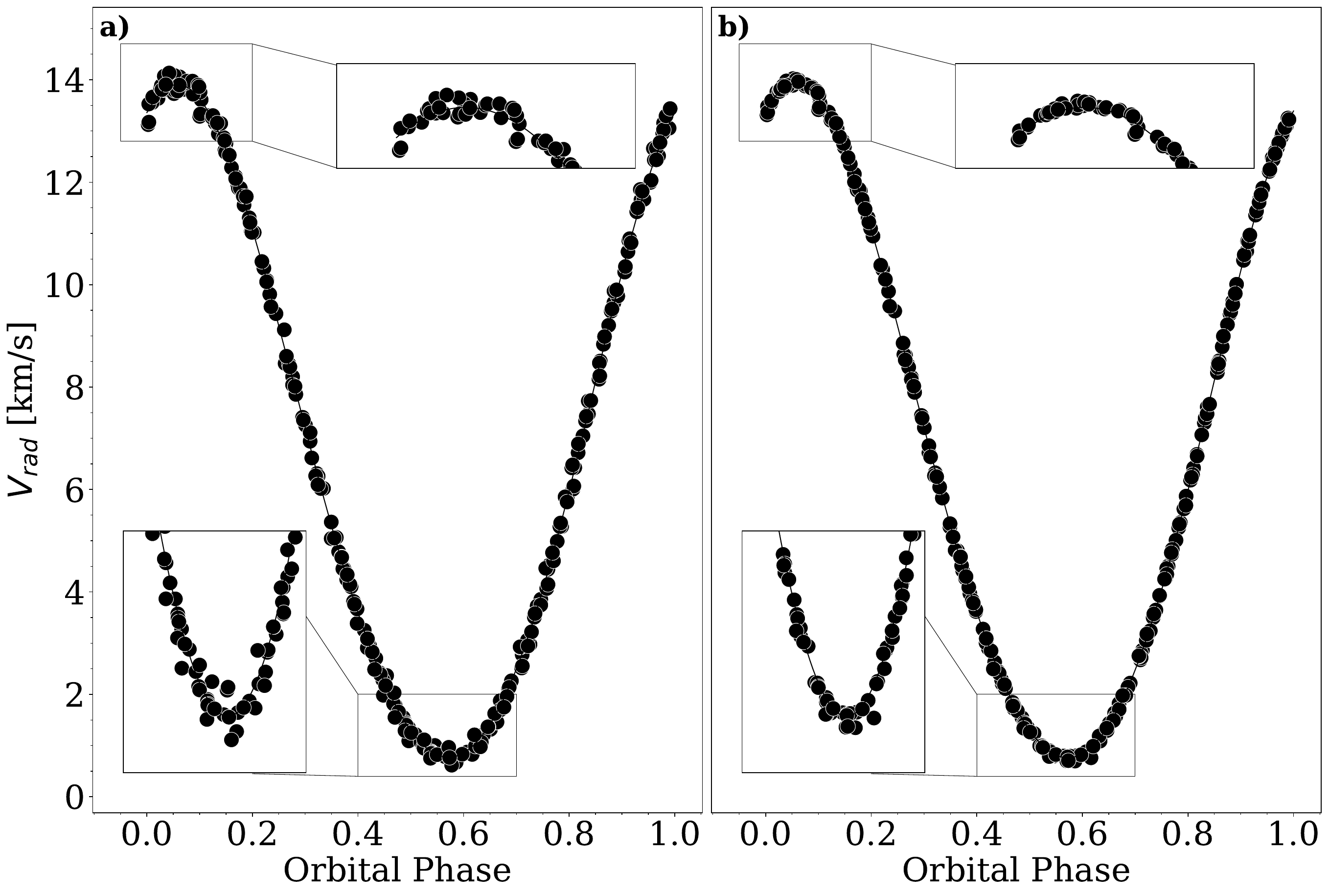}
    \caption{Phase folded RV curves for $\lambda$\,And. \emph{a.} Before removing the spot signal and \emph{b.}   after removing the spot signal. The maximum and minimum RV regions are enlarged for better viewing. The overall RMS decreased by a factor two. }
    \label{land_phases}
\end{figure}

\subsection{Spot correction} \label{Act_Corr}

A cool spot causes a spectral-line core to be net red-shifted when appearing on the approaching part of the rotating stellar hemisphere and a net blue-shift when on the receding part of the stellar hemisphere. The RV residuals versus phase then show a sinusoidal change centered on the starspot passage through the stellar central meridian. Once there is more than just a single spot, and viewed at moderate to low inclination, it is not simple anymore to imagine the spot's disk-integrated influence. Therefore, numerical simulation with spotted toy stars were carried out early on \citep{Saar&Donahue1997} and were perfected in order to quantify the effect for planet-hosting stars with ultra-precise RV observations \citep[e.g.,][]{Boisse2011, Meunier2023, Zhao2023}. In case of \lama, we have an active RS\,CVn star with super-sized spots compared to the Sun. This in turn allows us to actually reconstruct its surface spot distribution independently of the orbital solution by means of the Doppler-Imaging technique, which is usually not possible for any of the cool planet-host stars.

For the RV spot correction, we proceed as following: Once the observed RV curve was fit with an initial orbital solution, we used the observed minus computed ($O-C$) residuals as our spot tracer. The lower panel of  Fig.~\ref{rvs_res} shows these residuals for the $\approx$26 orbital revolutions (522\,d). Then we applied a LS periodogram which indicated the maximum power (amplitude squared) at a period of $P=54.41\,$d (Fig.~\ref{LS1}). Its most-likely error is $\pm$0.3\,d. The second strongest peak is at $P/2 \simeq 26.81\,$d, and the third strongest peak at $P/3 \simeq 17.75\,$d. The first period agrees very well with the $\approx$54\,d of the photometric period found by several authors in the past, \citep[e.g.,][]{Strassmeier1989, Henry1995, Parks2021}, and we interpret it as the rotational period. The existence of significant $P/2$ and $P/3$ harmonics indicate a stable two-spot configuration on the surface of \lama\ during our 522\,d observing window according to the models from \citet{Boisse2011}. By using three harmonics of a sinusoidal function, following \citet{Boisse2011}, a smoothed curve is fitted to the residuals. It is shown by the gray thick line in the lower panel of Fig.~\ref{rvs_res} along with the data points. Optimizing its fit to the RV curve based on a minimum $\chi^2$ approach results in a formally best fit with a slightly shorter period of 53.7\,d but with an uncertainty of 3.5\,d, still in good agreement with the more-precise LS period. This fit is then removed from the observed RVs. Its peak-to-valley RV amplitude is $\approx$500\,\ms.

\begin{table*}[ht!]
\caption{Orbital elements for $\lambda$\,And.}\label{lambda_and_orb}
\begin{flushleft}
\begin{tabular}{lllll}
\hline\hline
\noalign{\smallskip}
Parameter           & This paper,               & This paper,               & \citet{Walker1944}        &\citet{Burns1906}\\
                    & spot-uncorrected         & spot-corrected            &                           & \\
\noalign{\smallskip}\hline\noalign{\smallskip}
$P_{\rm orb}$ [d]   & \multicolumn{2}{c}{20.520394$\pm$0.000041}            & 20.5212$\pm$0.0003        & 20.546  \\
$K$ [\kms]          & 6.568$\pm$0.018           & 6.5815$\pm$0.0076         & 6.64$\pm$0.17             & 7.07$\pm$0.16 \\
$\gamma$ [\kms]     & 6.971$\pm$0.012           & 6.977$\pm$0.005           & 6.84$\pm$0.12             & 7.43$\pm$0.10  \\
$T_0$ (BJD)         & 2,459,696.53$\pm$0.13     & 2,459,696.372$\pm$0.056   & 2,429,202.39$\pm$0.48     & 2,416,683.46$\pm$0.39  \\
$e$                 & 0.0602$\pm$0.0028         & 0.0607$\pm$0.0012         & 0.040$\pm$0.024           & 0.086$\pm$0.018  \\
$\omega$ [deg]  & 339.5$\pm$2.3             & 336.8$\pm$1.0             & 313.6$\pm$8.9             & 301.0$\pm$7.6 \\
$a_1\sin i$ [Mkm] & 1.8499$\pm$0.0051      & 1.8536$\pm$0.0021         & 1.872                     & 1.990 \\
$f(m)$         & 0.000601$\pm$0.000005     & 0.000604$\pm$0.000002     & 0.0006                    & \dots \\
No. of obs.         & 265                       & 265                       & 27                        & 28 \\
Error of obs. of    & & & & \\
\ \ unit weight [\kms] & 0.134                   & 0.057                     & 0.45                      & 0.51 \\
\noalign{\smallskip}
\hline
\end{tabular}
\end{flushleft}
\end{table*}

\subsection{Orbital solution}

To model its orbital behavior, we treated \lama\ as a single-lined (SB1) binary and solved for the primary component using the general least-squares fitting algorithm from \texttt{scipy}\, \citep{SciPy2020-NMeth}. For solutions with non-zero eccentricity, we employed the prescription from \citet{dan:bur} to calculate the eccentric anomaly. A first orbit is determined from all available RVs including the ones tabulated by \citet{Walker1944} and \citet{Burns1906} -- data spanning 120\,yrs. It is used to fix the orbital period. By assuming that the fit is of good quality, we derive the element uncertainties (for all our solutions in this paper) by scaling the formal one-sigma errors from the covariance matrix using the measured $\chi^2$ values. $T_0$ is always a time of periastron. We note that we give orbital periods as observed and not corrected for the rest frame of the system.

Already our initial, spot-uncorrected orbit by far supersedes the previous best orbit from \citet{Walker1944}. However, there is no dramatic change of the individual elements, which hints towards the very high quality of the old Lick and Victoria data. Important is the consolidation of the non-circular orbit for \lama\ which we redetermine to an internal precision of 2\%. Our final orbit is then with the spot corrected RVs. As laid out in Sect.~\ref{Act_Corr}, the spot correction is achieved by subtracting the fit of the residual RVs from the observed RVs at the corresponding observing times. We then repeat the orbit determination and find our final elements. Table\,\ref{lambda_and_orb} lists the usual orbital elements and compares our orbits with those from \citet{Walker1944} and \citet{Burns1906}. The error of an observation of unit weight decreased from 134\,\ms\ for the spot-uncorrected orbit to 57\,\ms\ for the spot-corrected orbit. Elemental errors were lowered on average by a factor of three after the spot correction.

The computed RV curves are compared with our observed velocities in Fig.~\ref{land_phases}. Its panel \emph{a} shows the orbit with the spot-uncorrected RVs, while panel \emph{b} shows the final orbit with the spot-corrected RVs. Highlighted are the times of maximum and minimum velocity because these critically influence the orbital elements.

\section{Doppler imaging}\label{DI_section}

\subsection{Code summary and data input}

The stellar surface of \lama\ is reconstructed using the $i$MAP code \citep[e.g.,][]{carroll12}. We apply a multi-line inversion based on an average spectral line built from approximately 500 spectral lines with line depths larger than 60\%\ of the continuum. The main criterion for line selection is blending. Unaccounted blending is particularly worrisome for \lama\ because of its very small rotational line broadening of $\approx$7\,\kms. We input a line list with atomic data from the Vienna Atomic Line Database VALD-3 \citep{vald3} that covers wavelength range of 4800 - 5400 $\AA$ of PEPSI data. The averaging relies on a Singular Value Decomposition (SVD) algorithm and a bootstrap-permutation test to determine the dimension (rank) of the signal subspace; for details, we refer to \citet{carroll12}. The noise estimate for the SVD profiles are obtained from the error estimates provided by the bootstrap procedure. The weighted mean profiles have typical S/N of 20\,000 per pixel, compared to the $\approx$400-900 per pixel for an individual line.

The resolving power of our PEPSI spectra of 250\,000 (1.2\,\kms\ or 0.024\,\AA\ at 6000\,\AA), together with an average full width of the lines at continuum level of $2 \ (\lambda/c) \ v\sin i\approx0.3$\,\AA\ for \lama, enables an acceptable 12 resolution elements across the rotating stellar disk. According to the simulations of \citet{pis:weh}, five resolution elements are the minimum for successful DI. STELLA+SES spectra with $R=$55\,000 are therefore of too low a resolution and cannot be used for DI. 

For the computation of the local line profile, $i$MAP solves the radiative transfer per surface pixel for 72 depth points in a grid of tabulated Kurucz ATLAS-9 \citep{atlas9, atlas9grid} model atmospheres. The local line profiles are thus in 1D and in local thermodynamic equilibrium (LTE). The grid covers the temperatures between 3500\,K and 8000\,K in steps of 250\,K interpolated to gravity, metallicity, and microturbulence from the global synthesis fits (see Sect.~\ref{DI_input}). The stellar surface is thereby partitioned into 5$^\circ$$\times$5$^\circ$ segments (72$\times$36 pixels), resulting in a total of 2592 surface segments for the entire sphere. An inclination of 70$^\circ$ was adopted from the average of the inclinations taken from the literature and remained fixed for the inversion. We use an iteratively regularized Landweber method \citep[see][and references therein]{carroll12}. The Landweber iteration realizes a simple fixed-point iteration derived from minimizing the sum of the squared errors.

We neglect differential surface rotation during the inversion. Our observations were taken within a single stellar rotation; therefore, the differential rotation signal would be very weak.

\begin{figure}
    \includegraphics[width=8.5cm]{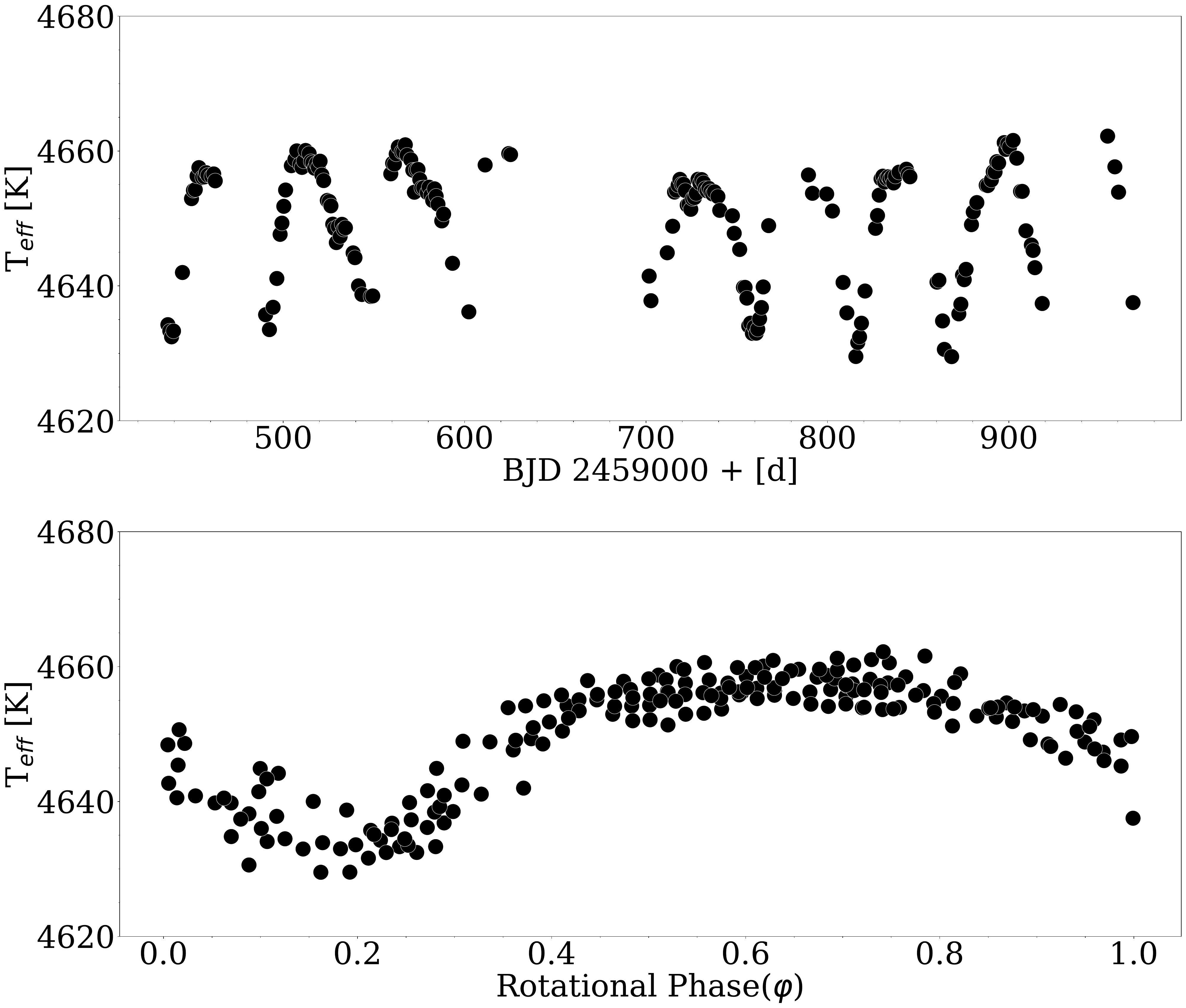}
    \caption{Effective temperature of $\lambda$\,And from ParSES. \textit{Upper panel:} Plot versus time throughout the 522\,d of observation in 2021-22. \textit{Lower panel:} Plot versus rotational phase. Folding is done with the rotational period from the RV jitter of 54.4\,d.}
    \label{surface_temp}
\end{figure}

\subsection{Stellar parameters}\label{DI_input}

Since we see no trace of a secondary star neither in the SVD line profiles, nor in the CCFs with the RV template, we proceed assuming all lines are from the primary star of the system. In addition, we adopt the 54.4-d period as the rotation period of the primary star which we inferred from the RV residuals.

Photospheric parameters are obtained by comparing the SES spectra with synthetic templates. We use the software Parameters-from-SES \citep[ParSES;][]{all04, parses} which utilizes a grid of pre-computed template spectra built with Turbospectrum \citep{turbo}, a line-list extracted again from the Vienna Atomic Line Database \citep[VALD-3;][]{vald3}, and a best-fit selection based on the minimum distance method with a non-linear simplex optimization \citep{all}. ParSES was applied to all STELLA spectra for the full 522-d coverage. The number of free parameters in the initial application of ParSES was five (effective temperature $T_{\rm eff}$, gravity $\log g$, metallicity [Fe/H], projected rotational velocity $v\sin i$, and microturbulence $\xi_t$). The averaged best-fit values are $T_{\rm eff}$=4655$\pm$47\,K, $\log g$=2.8$\pm$0.2 cm\,s$^{-2}$, [Fe/H]=--0.59$\pm$0.06, $v\sin i$=7.16$\pm$0.04\,\kms, and $\xi_t$=1.29$\pm$0.03\,\kms, where the errors are 1$\sigma$ rms values from the 184 individual spectra. We note that the most-likely spot-unaffected effective temperature is 4660\,K according to Fig.~\ref{surface_temp}. \lama\ had been known to be metal-poor with a listed 13 measurements in CDS/Simbad of around [Fe/H]$\approx$--0.5, a value in reasonable agreement with our fits. The range of recent rotational line-broadening determinations is between $v\sin i$=7.3~\kms\ \citep{Massarotti2008} and 6.5$\pm$0.3\,\kms\ \citep{Donati1995} but is not easy to constrain because of the similarly large macro turbulence and its expected enhancement in or around starspots \citep{Toner&Gray1988, Donati1995}. 

\begin{figure}
\includegraphics[angle=0,width=87mm]{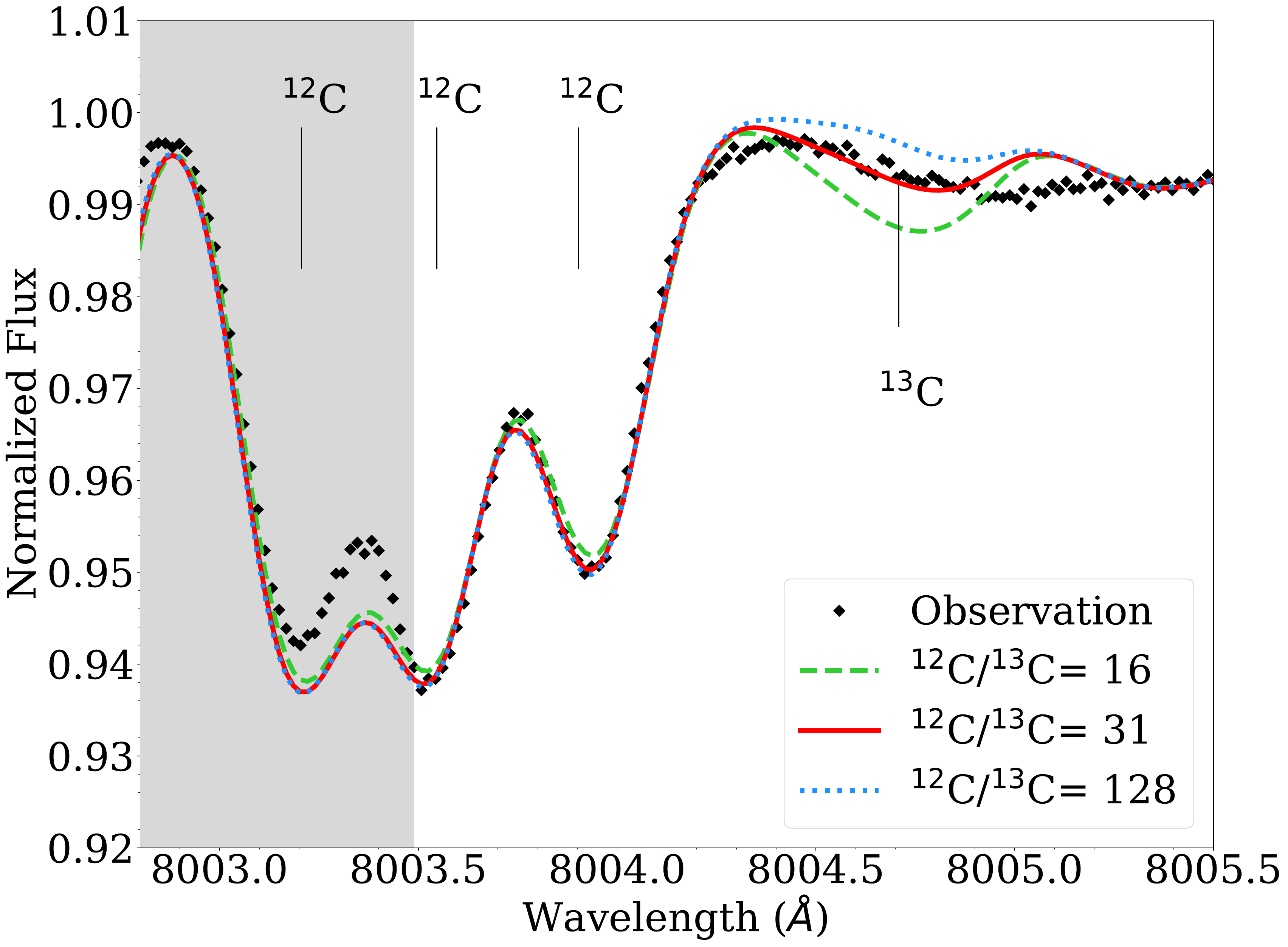}
\caption{Synthetic fits for the 8000-\AA\ CN region of \lama\,(gray shaded region is excluded for the fits). The diamond symbols are a phase-median spectrum corrected for telluric contamination. The lines are fits with three different $^{12}$C/$^{13}$C isotope ratios. The figure represents only one example of fits. Stellar parameters were kept fixed to ($T_{\rm eff}$, $\log g$, [Fe/H], $\xi_t$, $v\sin i$) = (4660, 2.80, --0.59, 2.00, 7.1) in the usual units using a relative nitrogen abundance $\Delta$A(N) of $-0.40$. In the case shown, we used the atomic and molecular line list of \citet{turbo}. The adopted best average from six parameter combinations is $^{12}$C/$^{13}$C=30$\pm$5 (the fit shown in the plot is with an isotope ratio of 31).}
 \label{F1213C}
\end{figure}

The upper panel of Fig.~\ref{surface_temp} shows the individual $T_{\rm eff}$'s from the ParSES fits as a function of BJD, and the lower panel as a function of rotational phase with the 54.4-d period obtained from the RV jitter. The $T_{\rm eff}$ time series allows a RV-independent period search. It yields a clear peak at $54.3\pm0.2$\,days and confirms the period from the RV jitter. Its error is estimated from the width of the resulting frequency peak because the covariance matrix is difficult to interpret in this case because every $T_{\rm eff}$ point has a different and not well-specified error. The phase-curve indicates a single-humped sinusoid with one broad maximum $T_{\rm eff}$ of 4660\,K at phase $\varphi$$\approx$0.7 and one relatively narrow minimum of just 30\,K cooler at 4630\,K at phase $\varphi$$\approx$0.2. The RMS of these fits is just 11.5\,K. The maximum temperature is in good agreement with the earlier spectroscopic analyses by \citet{Savanov&Berdyugina1994} who found $T_{\rm eff}$=4650\,K and $\log g$=3.0 but diverts from \citet{Tautvaisiene2010} who found $T_{\rm eff}$=4830\,K and $\log g$=2.8. We adopt the maximum effective temperature (4660\,K) from Fig.~\ref{surface_temp} as our starting value for the DI and keep the other photospheric parameters fixed to the averaged ParSES values.

\lama\ had been spatially resolved with the Center for High Angular Resolution Astronomy (CHARA) array and the Michigan Infra-Red Combiner (MIRC) instrument from observations in 2010 and 2011 by \citet{Parks2021}. They reported that the inclination of the star's rotation axis, $i$, from the two data sets in 2010 and 2011 is 75$\pm$5.0\degr\ and 66.4$\pm$8.0\degr, respectively, while \citet{Martinez2021} found an inclination of 85.6$\pm$2.3\degr\ from the same but combined data sets. \citet{Fionnagain2021} presented a Zeeman-Doppler image of \lama\ and found a best-fit inclination of 71$\pm$2\degr. Because of the small Doppler broadening, our spectral absorption lines do not present a good constraint for the inclination from fits to the line profiles. Therefore, we adopt a weighted average of above inclinations of 70\degr\ for our Doppler imagery.

The \textit{Gaia}~DR3 parallax of $38.57\pm0.12$\,mas \citep{DR3} refined \lama's distance to $25.924\pm0.079$\,pc as compared to $26.41 \pm 0.15$\,pc from the re-reduced Hipparcos data \citep{hip}. From the catalogue of \citet{Ducati2002}, we used the visual magnitude of $V=3.82$. By inherting the bolometric correction from \citet{Popper1980} we obtained the bolometric luminosity as 25.08 L$_\sun$, and with the interferometric radius $7.787\pm0.053$\,R$_\sun$ from \citet{Martinez2021}, we calculated the effective temperature to 4633\, K, which is close to the maximum temperature obtained by the ParSES spectrum fits (4660\, K). An independent radius estimate is obtained from the $v\sin i$, $i$, and $P_{\rm rot}$ relation (respectively taken as $7.0\pm0.5$\,\kms, $70\pm10\degr$\,and $54.4\pm0.3$\,d), yielding a radius of $8.0\pm0.4$~R$_\sun$ in good agreement with the interferometric radius. 

Of particular interest from an evolutionary point of view is the $^{12}$C/$^{13}$C carbon isotope ratio. Canonical stellar evolution models predict a core dredge-up that decreases the surface $^{12}$C/$^{13}$C and carbon/nitrogen (C/N) ratios over time. Specific dredge-up predictions for \lama\ suggested only a mild reduction in the $^{12}$C/$^{13}$C ratio to values $>30$ \citep[see the discussion][and references there in]{Drake2011}. While \citet{Drake2011} observed a C/N ratio for \lama\ that is perfectly consistent with the predictions, \citet{Savanov&Berdyugina1994} and \citet{Tautvaisiene2010} measured a surprisingly low carbon isotope ratio $^{12}$C/$^{13}$C of $20\pm5$\,and 14, respectively, that are not in agreement with the predictions nor with observations of low-metallicity, single, disk giants. 

Therefore, we also looked for the carbon isotope ratio in our very high resolution spectra. We first build a median spectrum out of the 40 individual PEPSI CD6 spectra and thereby correct for its (nightly variable) telluric contamination. This is possible because \lama\ is a SB1 binary with a precisely known orbit that easily allows identifying telluric lines from their fixed wavelengths. The pixels affected can thus be removed when building the median. This median spectrum has still a two-pixel resolution of 250\,000 and a S/N of 680 per pixel at 8000\,\AA. Program Turbospectrum \citep{turbo} is then employed with MARCS model atmospheres \citep{Marcs2008} and under the assumption of LTE to synthesize and fit the 8004-\AA\ wavelength region. One such fit is shown in Fig.~\ref{F1213C}. The synthesis approach follows our recent applications in \citet{vpnep} and for details we refer to its description. For \lama, we synthesize only a small part of the CN line regions around 8003.5\,\AA\ (due to $^{12}$C$^{14}$N) to 8005.9\,\AA\ (due to $^{13}$C$^{14}$N). We use again VALD3 complemented with the line lists of \citet{carl} and \citet{turbo} for CN and selected atomic lines. The fits result in a mixed bag of isotope ratios ranging from 21 to 35. Internal errors are very small but strongly depend on the level of the continuum. The true continuum for \lama\ may appear depressed by the (weak but unknown) molecular line content in the 8004-\AA\ region, and the fits thus depend on the adopted molecular line list. We conclude on a best-fit $^{12}$C/$^{13}$C isotope ratio of 30$\pm$5 for \lama\ placing it on the lower rim of dredge-up expectations of RGB members \citep{Drake2011}.  

\begin{figure*}[h!]
    \centering
    \includegraphics[width= \textwidth]{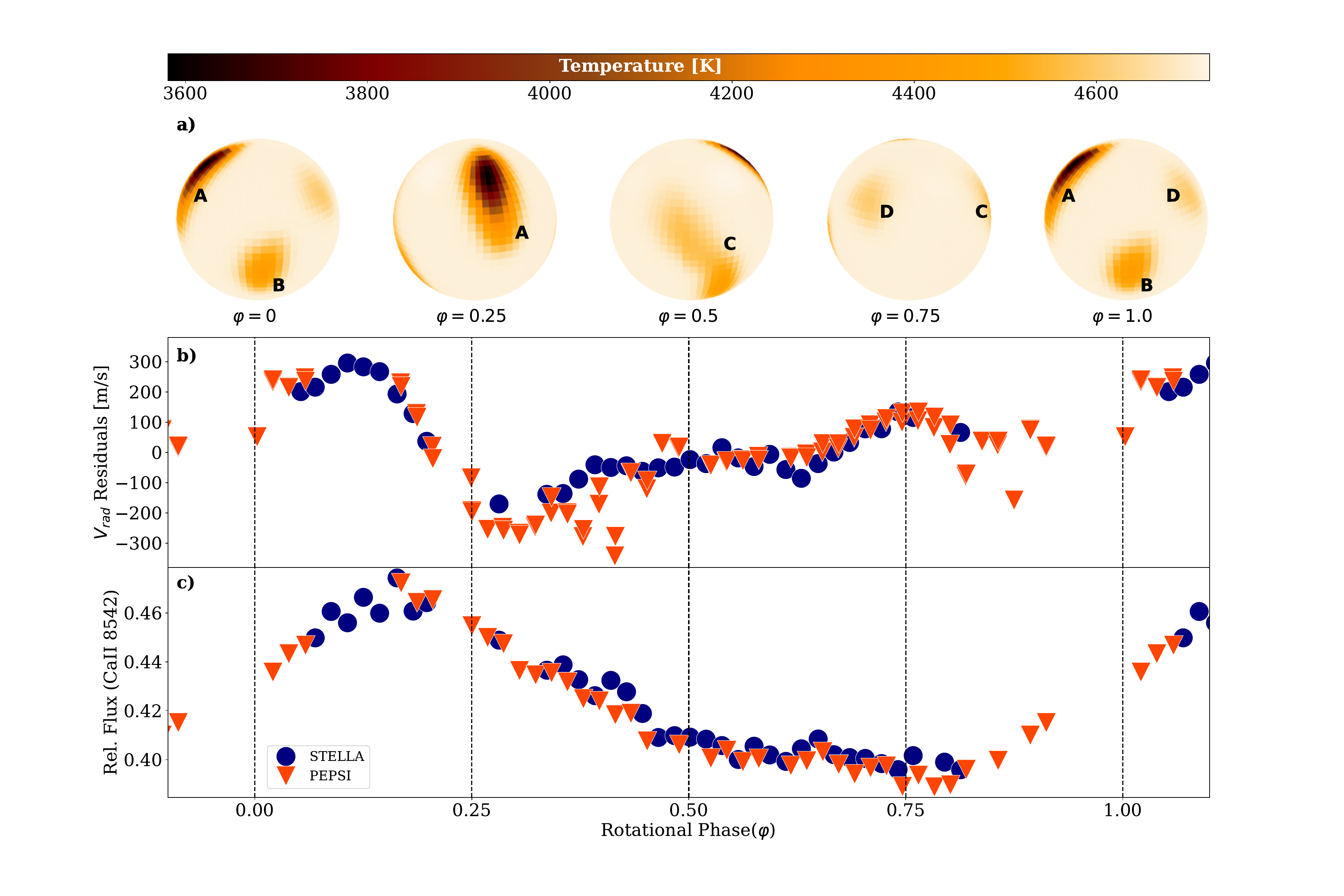}
    \caption{Doppler image, radial velocities, and chromospheric activity of \lama . Our Doppler image (panel \emph{a}) is compared with the RV residuals (panel \emph{b}) and the chromospheric \cairt\ (8542\,$\AA$) line core flux (panel \emph{c}). The Doppler image is from PEPSI spectra only and shows four spots, or spotted regions, identified with letters A--D. RVs correlate well with the location of the four cool spots and almost anti-correlate with the chromospheric \cairt\ emission. The shown RVs and activity data are from the same time-coverage than the Doppler image (May--June 2022).}
    \label{F_DI+RV}
\end{figure*}

\subsection{Doppler imaging results}\label{Results}

\begin{figure}[th!]
    \centering
    \includegraphics[width=9cm]{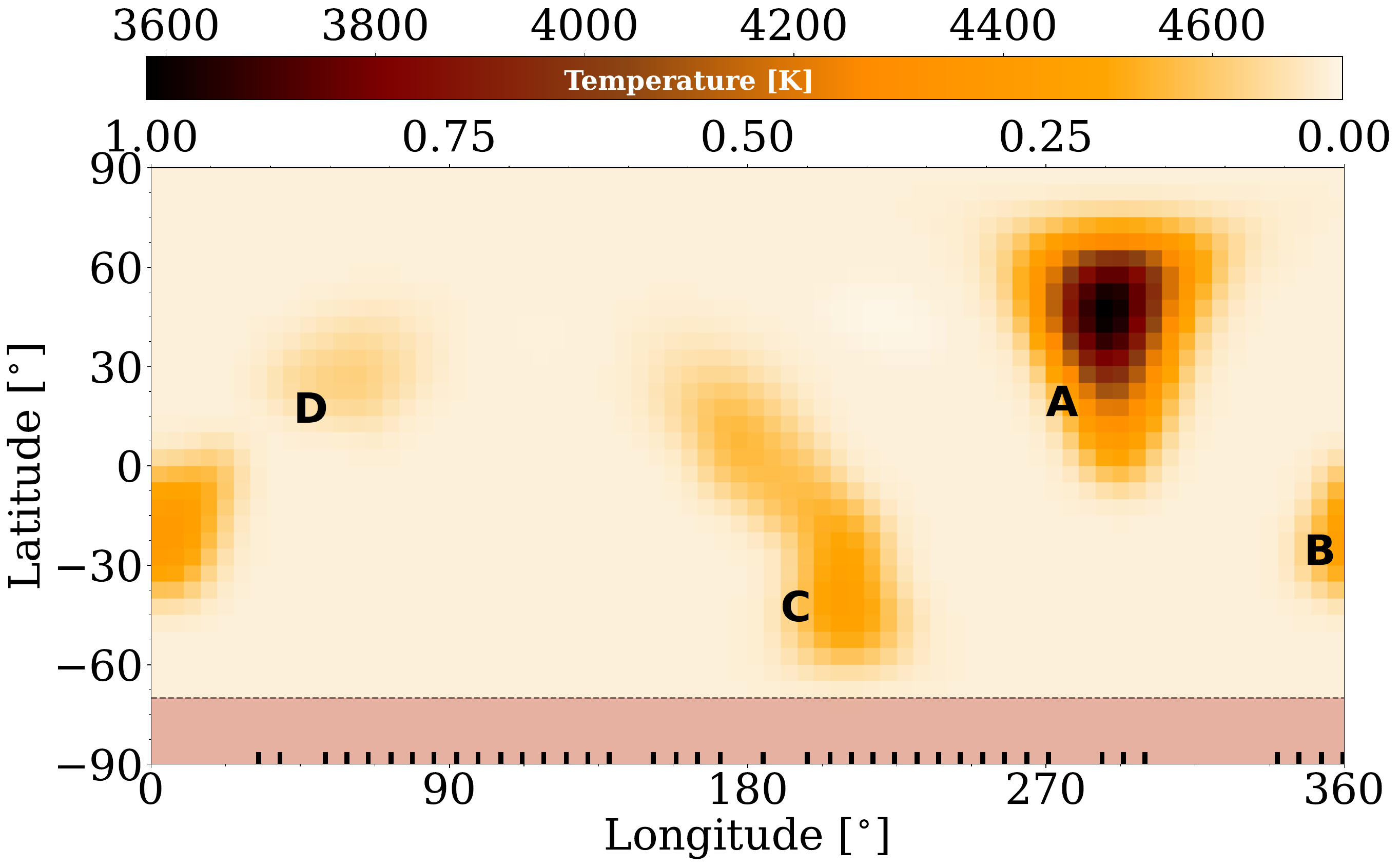}
    \caption{Mercator projection of the surface of \lama. The thick marks at the bottom indicate the phase of the observations. The red-shaded region below a latitude of $-70$\degr\ is invisible due to the inclination of the rotational axis of the star of 70\degr.}
    \label{Mercator_LA}
\end{figure}

Figure~\ref{F_DI+RV}a shows our Doppler image in orthographic projection at five equidistant rotational phases (first and last image are for the same phase). Figure~\ref{Mercator_LA} plots it in pseudo-Mercator projection. There are no polar spots during this one observing window in May-June 2022 nor are there bright faculae reconstructed. Our map indicates a four-spot surface structure at nearly all latitudes with one spot dominating. This spot, denoted spot~A, appears with its center at a latitude of around +40\degr. Its temperature contrast is up to $\simeq$1000\,K cooler than the photosphere in its central (dubbed umbral) area. Its two-temperature structure is a persistent feature of all our trial reconstructions but we nevertheless deem it uncertain due to the low line broadening of \lama. We note that such a sunspot-like morphology had been reconstructed also for the active solar-type single star EK~Dra \citep{ekdra} and could be potentially real. The other three spots, denoted B--D, appear comparably weaker and warmer than spot~A with temperature contrasts of $\simeq$300\,K cooler than the photosphere. Spot~D, the weakest of the four, is still consistently reconstructed but its existence is considered uncertain given the small $v\sin i$. Table~\ref{DI_spots} summarizes the parameters for the four features; location, temperature, and occupied area in units of the visible surface. The spot area is calculated by segmentation of the image with the given temperature values per pixel following \citet{Kovari2024}.

The umbral part of spot~A at a longitude of $\approx$290$^{\circ}$ occupies an area of 5\%\ of the hemisphere. However, the spot is more than three times larger if we also consider its penumbral part, which extends towards the equator. This penumbral part of spot~A is reconstructed with an average temperature contrast of $\approx$300\,K equal to the other spots. The umbral temperature for spot~A is consistent with what had been reconstructed previously from photometry \citep[by][$\approx$800--1000\,K]{Donati1995} and interferometry \citep[by][$\approx$800--1200\,K]{Parks2021} for other epochs. While light-curve temperature fitting relies on the rotationally-modulated color amplitude, mostly $B-V$, the interferometric temperature contrast is reconstructed implicitly from the H-band (1.65\,$\mu$m) visibilities and closure phases. Its errors are not available. Because surface parameters cross talk during image reconstruction, it is not clear how well the temperature contrast had been defined in the interferometric images. Nevertheless, the three different tracers (broad-band photometric, optically thin spectral line profiles, and H-band visibility) all agree around a spot temperature of 1000\,K cooler than the photosphere. 

Inside of such cool spots, the molecular bands are formed as shown in the simulations by, for example, \citet{Strassmeier2022}. TiO bandheads at 7055\,\AA\ or 7088\,\AA\,are typically used as an identifier of the cool spots on the stellar surface. For \lama,\,we do not detect the TiO bandheads, during the passage of the spot~A. The reason for this is that the surface temperature of \lama\,is just too high for such detection \citep[][]{Strassmeier2022}. On top of this, the spectral lines are blended by the telluric lines in that region. These two effects prevent this observation from occurring in our data.

The second spot, spot~B, at a longitude of 5$^{\circ}$, appears below the equator centered at $\approx$--20$^{\circ}$. It occupies an area of just 8\%\ and appears close to the central meridian at rotational phase zero. The third spot, spot~C, at a longitude of 240$^{\circ}$, appears also centered at a negative latitude similar to spot~B but stretches across the equator. In terms of occupied area of the visible hemisphere, spot~A and spot~C occupy the most area and both display approximately 20\% of areal coverage. Spot~D is the smallest among the resolved surface structures with an area of 5\%\ at a longitude near 60$^{\circ}$ and latitude around +35$^\circ$ with a temperature contrast of as small as 80~K, which is close to our likely detection limit of $\approx$50\,K depending on rotational phase.  

\begin{table}
  \caption{Spots on \lama\ in May-June 2022.}\label{DI_spots}
\begin{tabular}{l r r r r c}
\hline\hline\noalign{\smallskip}
Spot  & Long & Lat & $\Delta T_{\rm spot}$ & Area \\
ID    & (\degr)& (\degr) & (K) & (\% ) \\
\noalign{\smallskip}\hline\noalign{\smallskip}
A (umbra)       & 290 &  +40  & 1000 &  5 \\
A (penumbra)    &     &       &  300 &  12 \\
B     & 5   &  --20 & 280  &  8 \\
C     & 210 &  --40 & 230  &  20 \\
D     & 60  &  +30  & 80   &  5  \\
\noalign{\smallskip}\hline
\end{tabular}
\tablefoot{Longitudes and latitudes are given for the spot centers. The spot temperature difference is given for the coolest part of the spot. Only spot~A shows a dual-temperature structure dubbed umbra and penumbra. The spot area is given in per cent of the visible hemisphere.}
\end{table}

Figure~\ref{SVDs} in the Appendix compares the observed SVD profiles with the Doppler-imaging reconstructions. On the left hand side of the figure, we overplot the observed and calculated SVD profiles in different colors (red and black, respectively). The middle panel shows the $O-C$ residuals as a function of phase. As indicated, our Doppler data set comprises in total 40 observational phases for the single rotation of 54.4\,d. While this phase sampling is very good, we emphasize that the rather small $v\sin i$ of 7\,\kms\ is of the same order than the radial-tangential macro turbulence ($\approx$5\,\kms) which together lead to a line profile of width $\approx$9\,\kms. The impact of the Doppler effect is thus comparably small and spot bumps cannot be easily traced by eye anymore. Therefore, the best fit calculated from the SVD profiles is likely sensitive also to other intrinsic velocity fields of the star, for example, anisotropic macro turbulence.

This is believed to be seen during the first seven phases of our Doppler data set when the dominating spot~A is best visible. The fits to these seven line profiles are not as good as for the other phases and even display a systematic change with progressing phase. We recall that the basic stellar parameters for \lama\ were predetermined and then fixed during the inversion, notably $v\sin i$ and macro turbulence. It usually restricts us to employ globally averaged stellar parameters, which usually is not an issue with the typically large rotational line broadening for DI applications. However, in case of \lama, the line profiles are likely influenced by velocity fields from the vicinity of the dominating spot~A, maybe similar to the in- and outflow morphology of Sunspots \citep[e.g.,][]{Solanki2003}. While we cannot implicitly solve for such an anisotropy during the inversion due to the cross talk with the (already small) rotational Doppler effect, we may mitigate its impact by adapting different values of its global macro turbulence broadening for the (seven) phases in question. Such rotational phase-dependent line broadening had already been noticed by \citet{Donati1995} (refer to Fig.~10 in their paper). However, \citet{Donati1995} interpreted this activity-related change as extra Zeeman line broadening. 

\begin{figure}[th!]
    \centering
    \includegraphics[width=9cm]{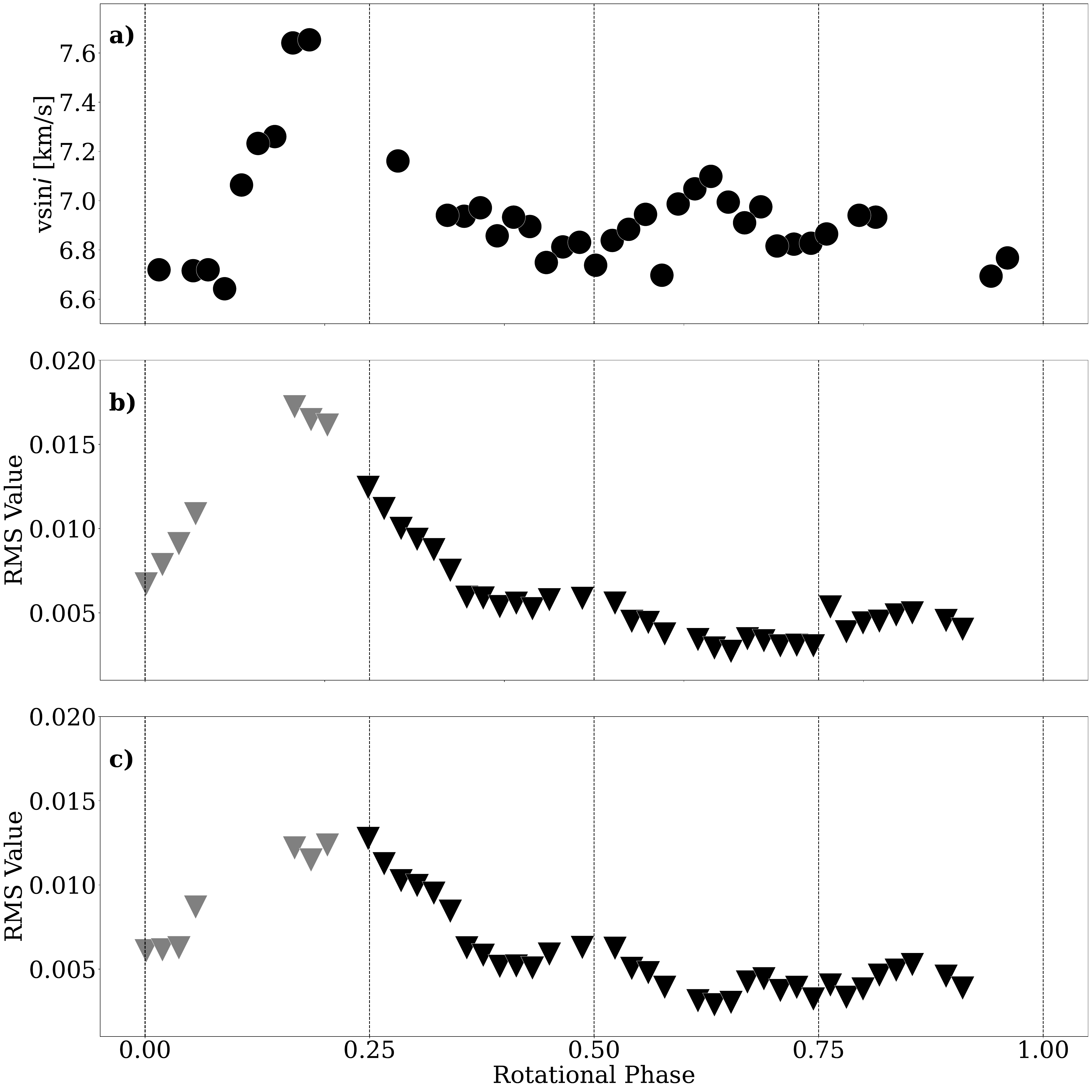}
    \caption{Line broadening and RMS residuals from Doppler imaging. Panel \emph{a}: Individual $v\sin i$ measures obtained from ParSES. Panel \emph{b}: RMS values of the SVD profiles from the initial Doppler imaging. Panel \emph{c}: Same as panel \emph{b} but after applying a variable macroturbulence broadening for the first seven phases. The grey markers indicate the seven phases where this was done.}
    \label{RMS_SVDs}
\end{figure}

Figure~\ref{RMS_SVDs} shows the RMS values for each line profile from the Doppler image and compares it with the independent $v\sin i$ measurements from ParSES (Fig.~\ref{RMS_SVDs}a). Figure~\ref{RMS_SVDs}b shows the RMS  from the initial Doppler-image fits with no change of macro  turbulence while Fig.~\ref{RMS_SVDs}c shows the improved RMS values including adaptive broadening for the first seven phases of the inversion. Because the synthetic SVD profiles were shallower than the observed ones for the first four phases ($\varphi$ = 0.001, 0.019, 0.038 and 0.056), we decreased the nominal macro turbulence by 1~\kms\ to 4~\kms\ for these four phases, that is less broadening. For the other three phases ($\varphi$ = 0.167, 0.185, 0.203) that show the largest RMS deviations, we increased the macro turbulence by 2~\kms\ to 7~\kms\, as the synthetic SVD profile cores appeared deeper than the observed ones, that is, we applied more broadening. This enabled lower RMS values for those phases as demonstrated in Fig.~\ref{RMS_SVDs}c. We emphasize that the variable line broadening is also seen from the SVD-independent ParSES $v\sin i$ fits, where the fitted values changed by up to 20\% during this epoch. It was these $v\sin i$ deviations that set our choice of correcting just the first seven phases (and not the following phases).   

We interpret the variable line broadening with an anisotropic turbulence component due to a moat-like flow around spot~A. As shown in Fig.~\ref{RMS_SVDs}a, the largest broadening (interpreted as $v\sin i$) is measured when spot~A is seen face-on in the direction of the observer. The $O-C$ discrepancy of the SVD-profile fits during DI is also largest when spot~A passes the central meridian. This suggests that the different broadening is indeed an activity-dependent feature likely related to the magnetic field of this one giant starspot. An anomalous effect is also seen in the RV residuals versus phase in Fig.~\ref{F_DI+RV}b, where the residuals show a change from positive to negative RVs at the time of the meridian passage of spot~A. 

The disk-integrated Doppler map converts to a rotational modulation of the photospheric temperature with an amplitude of just $\approx$35\,K. This relative temperature change agrees well with the $T_{\rm eff}$ change obtained from the independent ParSES spectrum-synthesis fits of the STELLA spectra in Fig.~\ref{surface_temp}, which gives $\approx$30\,K. However, we caution that our disk integration is only for guidance because we used SVD line profiles for the map which are built from line profiles spanning several thousands of Angstroems.

\section{Chromospheric activity and longitudinal magnetic field}\label{ca}

\subsection{Line-core fluxes in \cahk, \cairt, \Halpha, and \Hbeta}

Relative fluxes in a 1-\AA\ band centered on the central wavelength of the Ca\,{\sc ii} H (3933.7\,\AA) and K (3968.5\,\AA) lines were measured for all STELLA spectra (the PEPSI spectra did not cover this wavelength range). The same relative 1-\AA\ line-core fluxes were also measured for the Ca\,{\sc ii} infrared triplet (IRT) at 8498, 8542, and 8662\,\AA\ dubbed IRT-1, IRT-2, and IRT-3, respectively. For the latter wavelength range, we can make use of both STELLA and PEPSI spectra. Additionally, 1-\AA\ line-core fluxes were also determined for \Halpha\ (6562.8\,\AA) and \Hbeta\ (4861.3\,\AA). Both Balmer transitions appear in absorption for \lama. For the latter wavelength range, we can again make use of STELLA and PEPSI spectra. 

Our relative flux measurements are done by simply integrating the area between zero and the spectrum in a 1-\AA\ band. Its errors mostly depend on the spectrum normalization during the data-reduction process. For \lama, we matched all STELLA spectra to the continuum from the PEPSI spectra determined by a two-dimensional optimized extraction process included in its nominal data reduction. \Hbeta\ is blended with strong photospheric absorption lines, in particular from the \ion{Cr}{i} \ion{Fe}{i} pair around 4861.9\,\AA. We did not correct for this blending. 

Absolute emission-line fluxes are then obtained by multiplying the 1-\AA\ relative fluxes with the absolute continuum flux taken from the relations provided by \citet{Hall1996}. The logarithmic continuum fluxes for \lama's $B-V=1.01$ were thus fixed to 6.2060, 6.3911, and 6.4462\,\ergs\ per \AA\ for \cahk, \cairt, and \Halpha, respectively. No absolute calibration exists for \Hbeta . Phase-average values and rotational-modulation amplitudes are summarized in Table~\ref{T_fluxes}. Radiative losses are determined from the sum of the respective emission-line fluxes normalized to the bolometric luminosity and are 1.171$\times$10$^{-4}$ for \cahk\ and  1.328$\times$10$^{-4}$ for \cairt. Its errors are dominated by the error of $T_{\rm eff}$ (around 10\%) and the uncertainties of the absolute flux calibration (20--30\%). 

\subsection{Rotational modulation of line-core fluxes}

All seven chromospheric tracers show a consistent phase-dependent variability with maximum flux near phase $\varphi$ = 0.2 and a broad minimum around phase 0.7. Figure~\ref{Tracers} in the Appendix demonstrates the coherent rotational modulation for the entire time range of our data from January 2021 to July 2022. 

Figure~\ref{F_DI+RV}c compares the \cairt\ (8542\,$\AA$) emission-line flux with the Doppler image and the RV jitter. In this case only data from the time epoch of the one single stellar rotation in May--June 2022 were used. Its comparison emphasizes the spatial relation between the \cairt\ line emission and the cool spots. Spot~A is responsible for the maximum chromospheric emission-line flux, while the least-spotted appearance at phase 0.75 (see Fig.~\ref{F_DI+RV}a) coincides with the time of minimum chromospheric flux. We note that our DI had not reconstructed any features warmer than the photosphere that we could interpret as faculae. However, $i$MAP had repeatedly shown to be able to reconstruct such features, if present, from optically thin absorption lines, for example for the RS~CVn binary II~Peg \citep{Strassmeier2019}. For \lama, we attribute the lack of bright features in our imaging partly to the low rotational line broadening and its vulnerability to anisotropic macro turbulence (and other velocity fields). The non-detection thus remains preliminary. 

\begin{table}
\caption{Logarithmic absolute emission-line fluxes in \ergs\ for \lama.}\label{T_fluxes}
\begin{tabular}{lllll}
\hline\hline\noalign{\smallskip}
Bandpass  & Continuum       & Average    & Variability & Fit   \\
          & flux            & line flux  & amplitude   & rms   \\
          & (per \AA )      &            &             &       \\
\noalign{\smallskip}\hline\noalign{\smallskip}
Ca\,{\sc ii} 8498 & 6.3911 & 6.154 & 0.075 & 0.012\\
Ca\,{\sc ii} 8542 & 6.3911 & 6.024 & 0.070 & 0.011\\
Ca\,{\sc ii} 8662 & 6.3911 & 6.029 & 0.065 & 0.010\\
Ca\,{\sc ii} H    & 6.2060 & 6.135 & 0.3   & 0.040\\
Ca\,{\sc ii} K    & 6.2060 & 6.247 & 0.3   & 0.051\\
\Halpha           & 6.4462 & 6.026 & 0.035 & 0.012\\
\Hbeta            & \dots  & \dots & 0.010 & 0.005\\
\noalign{\smallskip}\hline
\end{tabular}
\tablefoot{There is no empirical continuum flux calibration for \Hbeta. Variability amplitude and fit rms as   relative fluxes in units of the continuum ($=1$), see Appendix Fig.~\ref{Tracers}.}
\end{table}

\subsection{Snapshot longitudinal magnetic field}

\begin{figure}[tb]
\hspace{6mm} {\bf a)} \\
    \includegraphics[width=87mm]{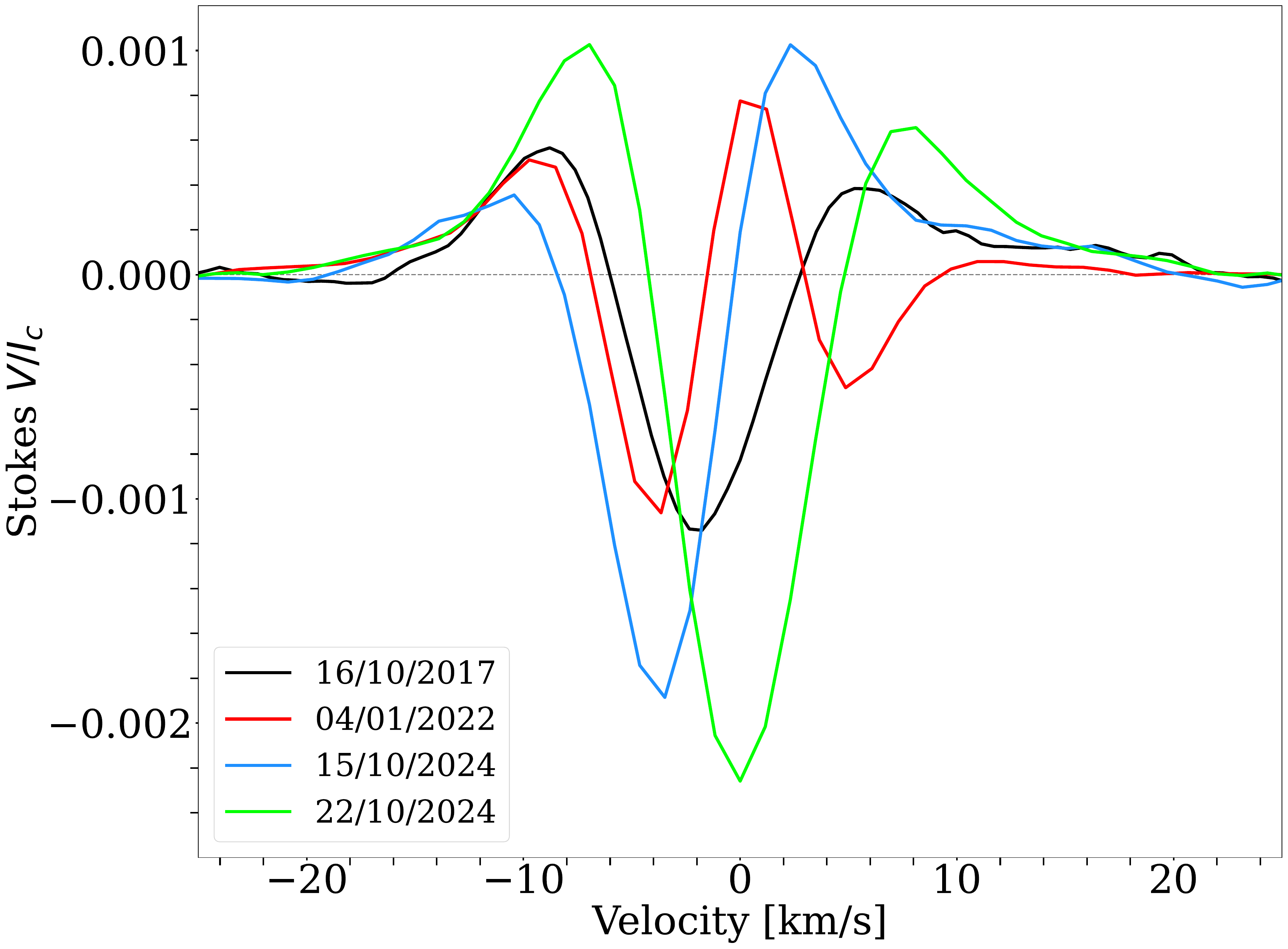}
    
\hspace{6mm} {\bf b)} \\
    \includegraphics[width=87mm]{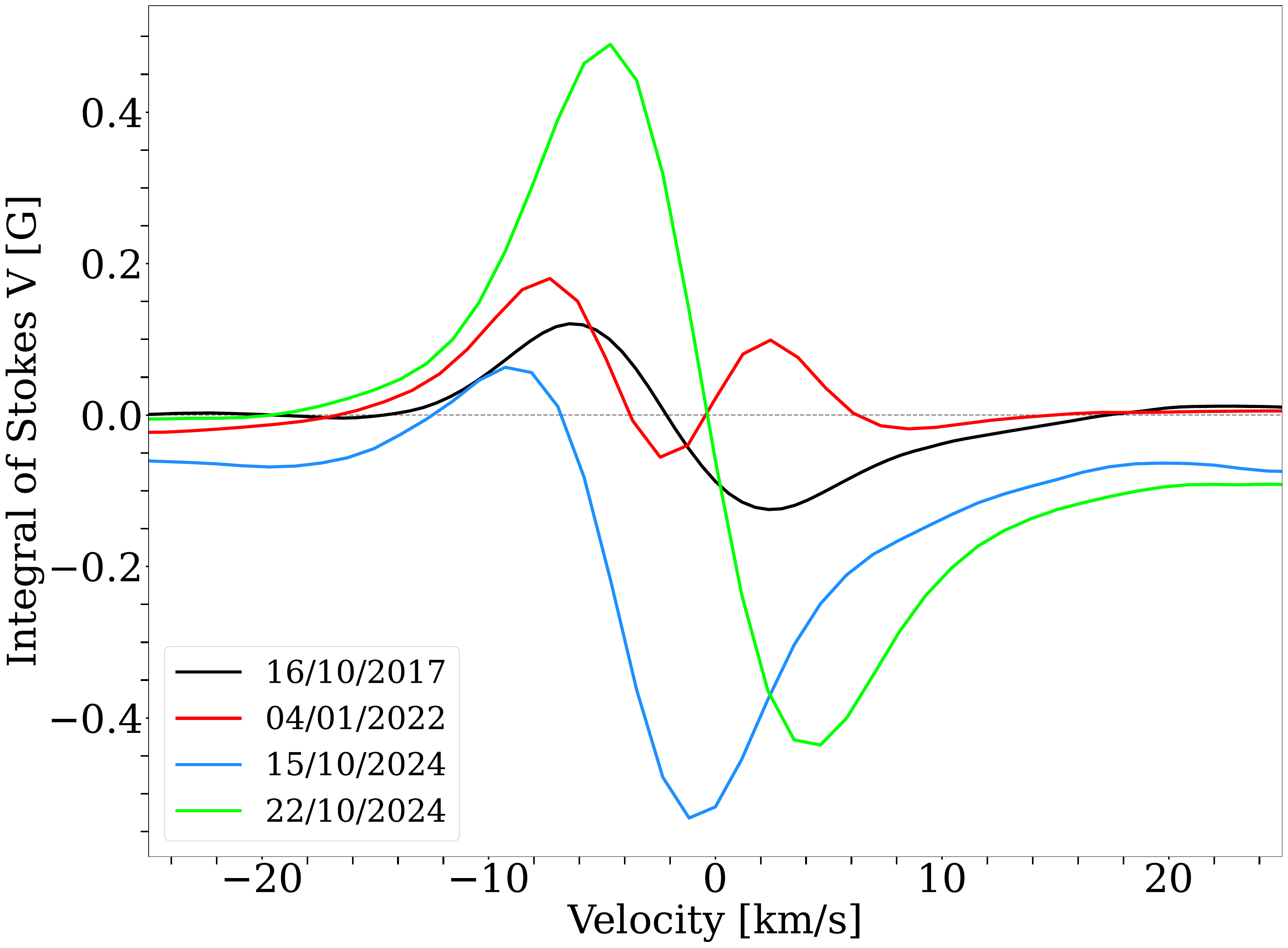}
    \caption{Stokes-V profiles for \lama. \emph{a)} PEPSI LSD line profiles in units of normalized intensity. \emph{b)} Integrated Stokes-V profiles in units of Gauss. Shown are our two snapshot observations from October 2017 (black), January 2022 (red) and two measurements from October 2024 (blue and and green). The observation contemporaneous to our Doppler image (May-June 2022) shows a very complex integrated field structure with two large bi-polar regions on the surface visible at the same time.}
    \label{StokesV}
\end{figure}

Magnetic field measurements are obtained from four high-resolution PEPSI Stokes-V spectra from four different days in years 2017, 2022, and 2024. The spectra from different cross dispersers (see Table \ref{T_magfield}) were line averaged with a Least Squares Deconvolution (LSD) technique by using the VALD-3 line list, in a similar way that, for example \citet{Kochukhov2010} has applied. These LSD line profiles are shown in Fig.~\ref{StokesV}a. 

We obtain a disk-integrated mean longitudinal magnetic field of $-0.51 \pm 0.26$\,G for October 2017 and $+0.54 \pm 0.20$\,G for January 2022. In October 2024, we found higher values of $-2.72 \pm 0.34$\,G and $+2.66 \pm 0.35$\,G, in all cases following the notions in \citet{Sol&Ste1984}. The integrated absolute (unsigned) values, |$B$|, are 4.0\,G and 2.0\,G for 2017 and 2022, respectively, and 10.2\,G and 11.4\,G for 2024. We used an equivalent wavelength of 5840\,\AA\ and a Land\'e factor of 1.22 while calculating the LSD profiles. The signed mean values for 2017 and 2022 are only about twice as large as the strongest Sun-as-a-star values that were recently measured by \citet{sdipol}. However, the Stokes-V spectrum of \lama\ from Jan.~2022 exhibits a more complex LSD profile than the Sun-as-a-star with two minima and two maxima with a full amplitude of $1.5\times 10^{-3}$, which is ten times larger than for the Sun-as-a-star. 

Figure~\ref{StokesV}b shows the Stokes-V integrated magnetic-field profiles following the procedure of \citet{Sol&Ste1984}. It indicates two (strong) positive-polarity regions and at least two (weaker) negative-polarity regions being present on the projected stellar disk at the same time in Jan.~2022. The Stokes-V profile from Oct.~2017 appears with only one central minimum and two adjacent maxima and looks thus similar to profiles from 2016 in \citet{Fionnagain2021}. Its integrated profile suggests only a single positive and a single negative polarity on the visible disk. The Jan.~2022 profile, that is close in time to our Doppler image, shows for the two maxima a velocity difference of 10.7\,\kms, while the two minima show a similar separation of 9.2\,\kms. This indicates that we have likely two active regions but each of a bipolar structure. Such a complex profile was not among the ones used for the ZDI map in \citet{Fionnagain2021} from 2016. We note that the one big spot~A in our contemporaneous Doppler image was not visible during neither of our two magnetic-field measurements. 

\begin{figure}[th!]
    \centering
    \includegraphics[width=87mm]{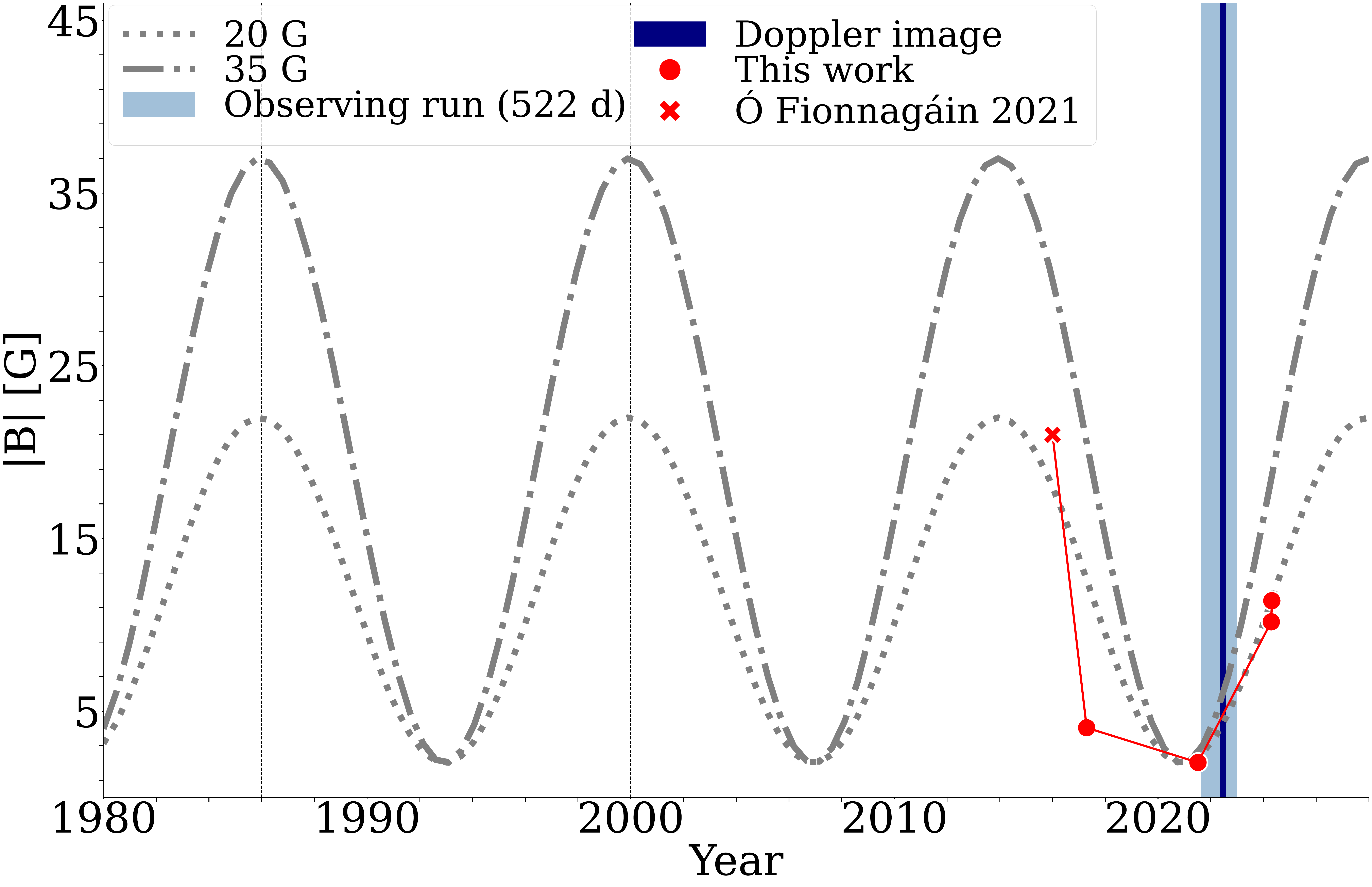}
    \caption{Long-term magnetic field measurements (shown in red symbols) and possible cyclic changes. A likely cycle period of 14 years from photometry \citep{Mirtorabi2003} from 1976--2002 was adopted. They identified two activity peaks in 1986 and 2000 (indicated with two vertical lines). Two versions of an assumed sinusoidal variation with a maximum amplitude of |$B$| of 20 and 35\,G are shown. The respective minimum value was set to 2\,G which is the minimum of the measured absolute magnetic flux in 2022. Our field measurements coincided with a time of activity minimum.} 
    \label{cyc14}
\end{figure}

Regarding the unsigned mean field, \citet{sdipol} measured a disk-averaged line-of-sight magnetic field of the Sun-as-a-star of +0.37\,G with PEPSI that related to an unsigned mean field of $\approx$13\,G as deduced from a nearly simultaneous full-disk line-of-sight SDO\footnote{Solar Dynamics Observatory; https://science.nasa.gov/mission/sdo} magnetogram. Assuming such solar scaling for \lama, we may expect a true unsigned large-scale (mean) field on the surface of \lama\ of as large as $\approx$100\,G. Figure~\ref{cyc14}  indicates a |$B$| trend paralleling the predicted photometric cycle variability with a minimum near the year 2021.

\begin{table}
\caption{Magnetic field measurements for \lama.}\label{T_magfield}
\begin{tabular}{llllll}
\hline\hline\noalign{\smallskip}
BJD            &PEPSI      &Phase        & \multicolumn{2}{c}{$B_{\rm long}$}      &|$B$|           \\
               &CD\#        & ($\varphi$)  & \multicolumn{2}{c}{[G]}            & [G]          \\
\noalign{\smallskip}\hline\noalign{\smallskip}
2,458,042.7426   &3,4,5   & 0.608       & \multicolumn{2}{c}{$-0.52 \pm 0.26$}   & 4.04   \\
2,459,583.5896   &3,4,5,6 & 0.927       & \multicolumn{2}{c}{$+0.54  \pm 0.20$}   & 2.03   \\
2,460,598.6609   &3,5     & 0.583       & \multicolumn{2}{c}{$-2.72 \pm 0.34$}   & 10.17  \\
2,460,605.9344   &3,5     & 0.717       & \multicolumn{2}{c}{$+2.66  \pm 0.35$}   & 11.39  \\
\noalign{\smallskip}\hline
\end{tabular}
\end{table}

\section{Discussion and summary}\label{Discussion}

\subsection{Rotational modulation of orbital RV curves}\label{Spots_RV}

Numerical simulations of spot-related RV modulation (loosely referred to as RV jitter) by \citet{Boisse2011} had shown that a two-spot structure on a rotating star leads to three peaks in the Lomb-Scargle periodogram. The highest peak appears at the rotational period of the star and the others at half and one-third of this period. \citet{Boisse2011} also showed that the modulation has the shape of a single-sinusoidal curve when the longitude difference of two spots is 60$^{\circ}$, and a double-sinusoidal curve if that difference is 120$^{\circ}$, with RV amplitudes of $\approx$25 and 50~\ms, respectively. Spot size was 1\% of the visible surface. Our observed spot structure for \lama\ is more complicated than such a model but the average longitudinal difference between the consecutive spots~A, B, and C is about $\sim75^{\circ}$, that is in between the two-spot structure in the simulations of \citet{Boisse2011}. Our observations of \lama\ confirm this with even four spots on the stellar surface (see Fig.~\ref{LS1}). 

The effects of (bright) faculae and (dark) spots on RV residuals were simulated and discussed most recently by \citet{Zhao2023}. Depending on which component is dominating, the shape of the respective RV curve is different while the combined effects spots+faculae almost annihilate each other and suppress the total residual RV amplitude. For a K2 star with an effective temperature of 5100\,K, and a 1\% areal coverage for both spots and faculae,  \citet{Zhao2023} predict residual RV amplitudes between 1--4\,\ms\ for sunspot-sized starspots. In Fig.~\ref{F_DI+RV}b, the peak-to-valley RV amplitude for \lama\ is $\approx$500\,\ms. Even if we scale the simulated spot size up by a factor ten to match the \lama\ observations, the predicted RV amplitudes are still a factor ten below the observation. 

The shape of the RV curve of \lama\ between rotational phases $\varphi$=0.3 to 0.5, where spot~C is entering the line-of-sight, changes more irregularly than during the passage of the dominating spot~A at around phase 0.2. A similar irregular change appears after half a rotation when spots~B and D are visible. This suggests that the spots on \lama\ are always accompanied by a dynamic surrounding photosphere. Albeit faculae are not identified in our (photospheric) Doppler image, we see the clear rotational modulation of chromospheric tracers in phase with the spot locations. It suggests a large-scale chromospheric inhomogeneity, likely due to individual plages right above the photospheric spots. However, the large residual RV amplitude of $\approx$500\,\ms\ indicates that it is more likely that local velocity fields in and around spots are the main contributor to the RV jitter. The simulated RV curves from a combination of solar-like faculae/plages and spots, together with the associated suppression of the convective blue shift, is likely still present but approximately a factor ten weaker than the spot-related surface dynamics. 

\subsection{Impact of spots on RV-based binary orbits}

Removing the rotational modulation from the orbital RV curve of \lama\ lowered the effective error of an observation of unit weight from 134\,\ms\ to 57\,\ms. Element errors were lowered on average by a factor two to three after the spot correction, which enables us to have a precise mass function of just 0.000604$\pm$0.000002. However, what is important for \lama\ is that the new accuracy possibly allows a glimpse into more subtle surface forces of orbital-rotational synchronization. 

The very small but still non-zero eccentricity (by 50$\sigma$) seems most readily be explained by (normal) tidal effects subsequent to the star formation \citep[e.g.,][]{Hut1981}. The fact that the relatively slow and sub-synchronous rotation of \lama\  is still accompanied by strong magnetic activity suggests the excess orbital energy (over a circular orbit) is likely used in \lama\ for extra stirring the convective surface layers. While speculative it would qualitatively explain the non-solar large-scale magnetic activity seen in the ZDI of \citet{Fionnagain2021} as well as  our observation of spot-related enhanced surface dynamics.     

\subsection{About long-term orbital period variations}

The orbital period of 20.520394$\pm$0.000041\,d in Table~\ref{lambda_and_orb} had been reconciled from all existing RV data, covering basically 120~yrs. When we use only the STELLA and PEPSI RVs, covering 522\,d, the best-fit  orbital period is shorter by 23$\sigma$, 20.5191$\pm$0.0012\,d. As \citet{Applegate1992} suggested, a long-term modulation of the orbital period of an active binary, $\Delta P/P$, could amount to $\sim$10$^{-5}$ due to the activity-dependent distortion of the gravitational quadrupole moment. Assuming above period difference of 23$\sigma$ is real, it could be interpreted as an orbital period modulation of $\Delta P/P \sim 6.3 \times 10^{-5}$. However, we note that the earlier period determinations by \citet{Walker1944} and \citet{Burns1906} are not precise enough for a comparison at the $10^{-5}$ level.

In order to check for a possible third body bound to the system, we analyzed the RV residuals after spot correction for remaining periodicities. These RV residuals are shown together with its LS periodogram in the Appendix in Fig.~\ref{RV_Res2}. We did not detect any indication for a third body in the time window given, neither a periodicity nor a trend, which could have been a hint for the grossly asynchronous behaviour of the orbital and the rotational dynamics. Our activity corrected orbital rms error of 57\,\ms\ is comparable to the residual 48\,\ms\ error for V830~Tau obtained by \citet{Donati2016} after filtering its activity and exoplanet signals. However, it is principally possible that the residual RV variations are still camouflaged by the imperfect subtraction of the dynamic stellar activity by using a static spot model. Repeated orbital observations and determination of the orbital elements will eventually help draw a conclusion about the existence of a third body.

\subsection{Photospheric and chromospheric surface activity}

\cahk\ and \Halpha\ variability of \lama\ had already been found to correlate well with its rotation period \citep{Bal:Dup1982}. Our data now show this also for the calcium infrared triplet and \Hbeta. While the absolute emission line fluxes of \lama\ are comparable in strength to those for the active Sun-as-a-star, despite the two star's $\approx$1000-K surface temperature difference, the chromospheric radiative losses in units of the respective bolometric luminosity are $\approx$50\% higher than for the Sun. 

Our Doppler map, together with the paralleled modulation of the emission-line fluxes, suggests a tight spatial relation of the large photospheric spots with the regions of chromospheric line emission. The latter remain disk unresolved from our data but are likely from a combination of a global chromospheric network and localized plages, as is the case on the Sun. Both observations, the amount of chromospheric radiative losses and their correlation with the morphological features (i.e., spots) on \lama, are explainable by scaled solar activity. 

Similar may apply to macro turbulence for \lama. We recall that our best-fit values for the rotational line broadening and for macro turbulence line broadening is 7.0\,\kms\ and 5.0\,\kms, respectively. This means that our DI line-profile fits are sensitive to macro turbulence and its radial-tangential component structure since its vector sum is almost as large as the rotational broadening. Our adopted macro turbulence is still uncertain by $\pm$1\,\kms, which is an annoying 20\% because it depends on the spot distribution on the stellar surface. This makes it more difficult to reach a near-perfect fit between the observed and the computed SVD line profiles, which is already demanding given the very high S/N. Because we input macro turbulence explicitly by adding a fixed broadening to the local line profiles prior to inversion, we usually neglect the effects of, for example,  latitude dependent turbulence that may (or may not) show up with different radial tangential components at different latitudes.

The observed phase-dependent line broadening together with our DI line-profile fits indicate that the global macro turbulence must be anisotropic, in case of \lama\ being basically related to spot~A. We interpret this anisotropy with a moat flow \citep[e.g.,][]{Toner&Gray1988} around spot~A. Our DI reconstructs spot~A with a large penumbra in the form of lowered temperature. However, we emphasize that our DI cannot take into account local surface velocity fields. The inversion would interpret profile deviations always as a temperature difference (if the spot-bump amplitude allows this).  Otherwise, it would show up as a phase-dependent lowered fit quality. This is the case during the meridian passage of spot~A as shown in Fig.~\ref{RMS_SVDs}b. The line profiles for these phases required a change of broadening of order 1--2\,\kms\ with respect to the nominal macro turbulence.  

\citet{Fionnagain2021} showed with their ZDI map from 2016 that the strongest magnetic activity came entirely from equatorial latitudes within approximately $\pm$40$^{\circ}$. The resolved spots in our Doppler image appear between latitudes --40$^{\circ}$ to +70$^{\circ}$ including the range seen in the ZDI. Despite global magnetic activity usually being best seen from the radial field component, the \lama\,ZDI showed an equally strong azimuthal component originating, similar to the radial component, solely from equatorial regions but being of only positive polarity. However, it appears that the stellar latitudes in the ZDI by \citet{Fionnagain2021} are not well constrained, if at all, as indicated by their unsatisfactory reduced-chi-square fit to the Stokes-V profiles. The authors attributed this to unaccounted surface differential rotation and intrinsic field changes. 

Our Doppler image shows a total spot coverage of between 5\% to 20\% of the visible surface. This range is comparable to the findings of \citet{Parks2021} and \citet{Martinez2021} from interferometric imaging. The detailed surface distribution in our Doppler image from May-June 2022 is different to the interferometric images though, which were obtained in 2010 (11 nights spanning 39 days) and 2011 (6 nights spanning 22 days). In both papers, \citet{Parks2021} and \citet{Martinez2021}, there are multiple spots reconstructed within latitudes of 0--30$^{\circ}$ (maybe with one exception of a weaker spot at approximately  --30$^{\circ}$ in 2011). Spot temperatures of as cool as 1200\,K from H-band interferometry appear generally comparable with DI but are formally lower by 200\,K than even for our coolest spot~A. In both the 2010 and 2011 epochs, \citet{Martinez2021} indicated that the coolest spots appeared near +20--30$^{\circ}$ latitude. Our Doppler image shows the dominating spot at slightly higher latitude of +40$^{\circ}$. In the opposite hemisphere, below the equator, our Doppler image resolves two more spots at a latitude of --20$^{\circ}$ and --40$^{\circ}$ (spots B and C, respectively) comparable to the one spot at --30$^{\circ}$ in the interferometric image from 2011. Besides, the interferometric images reconstructed  no dominant spot concentration in the opposite hemisphere below the equator. This kind of north-south asymmetry had been interpreted by \citet{Martinez2021} as evidence of a non-solar dynamo. 

\subsection{Surface magnetic field and magnetic braking}

ZDI of \lama\ by \citet{Fionnagain2021} revealed a strong large-scale magnetic field with an average unsigned field strength of 21\,G (in 2016), with most of the magnetic energy in the radial component. Its low-latitude azimuthal field component even formed an unipolar ring around the entire stellar surface. Our (snapshot) measurements yield a mean longitudinal field $B_{\rm l}$=2.6\,G in 2022 (and 3.9\,G in 2017), compared to the maximum of the time series in 2016 of 13.7\,G \citep{Fionnagain2021}. The new high-resolution Stokes-V/I profiles from PEPSI hint toward a more complex surface field geometry than previously observed. 

The currently detected unsigned mean magnetic field of $\approx$10\,G supports the hypothesis that magnetic braking had and still is working in the \lama\ system, and that it likely had outperformed the counteracting tidal forces all along as had already been emphasized by \citet{Donati1995} and modelled by \citet{Garraffo2015, Garraffo2018}. As \citet{Donati1995} calculated, a mean magnetic field should be greater than $\approx$30\,G in order to make magnetic braking faster than synchronization. On the other hand, \citet{Garraffo2015} showed numerically that for the most efficient angular momentum loss, the spot locations should be distributed in lower latitudes. That condition seems to be  confirmed by our Doppler image. The current best explanation for the asynchronous rotation of \lama\ is thus magnetic braking. Unfortunately, our current observations are not fully conclusive and do not yet constrain a specific value for the fraction of open versus closed magnetic field lines and its associated mass loss.   
 
Important in the connection with the system's asynchronism may be that the orbit of \lama\ has a remaining non-zero eccentricity ($e=0.0607\pm0.0012$), meaning that perfect orbital circularisation has not yet been achieved. Tidal theory predicts that synchronisation should have occurred before circularisation \citep[e.g.,][]{Hut1981}. While the SB1 mass function in Table~\ref{lambda_and_orb} is of very high precision (and likely accuracy) we still need to know the primary mass from an independent source in order to determine the secondary-star mass from the mass function. This can be done from the physical radius and gravity of the primary. Our best nominal values from Sect.~\ref{DI_input} suggest a mass of $1.4^{+0.8}_{-0.5}$\,M$_\sun$, where the errors are dominated by the gravity error. While there are consistent gravity determinations, albeit with large errors, basically ranging between $\log g$ of 2.5 and 3.1 (cm\,s$^{-2}$), the inferred radii range from $6.40\pm0.16$\,R$_\sun$ based on $V$, $T_{\rm eff}$, and the \textit{Gaia} DR3 parallax to $8.0\pm0.4$\,R$_\sun$ based on $v\sin i$, $i$, and $P_{\rm rot}$. If we adopt $\log g=2.8\pm0.2$ and $R=6.40$\,R$_\sun$, we obtain a primary mass of only $0.94\pm0.32$\,M$_\sun$. If we adopt $R=8.0$\,R$_\sun$ (same $\log g$ value), we obtain a mass of $1.47\pm0.44$\,M$_\sun$. These values convert to secondary masses in the range 0.086 to 0.115\,M$_\sun$ and mass ratios $q=M_2/M_1$ of 0.09 to 0.08, respectively, suggesting a companion that is barely above the hydrogen-burning limit of $\approx0.075$\,M$_\sun$. The unseen secondary of \lama\ is thus most likely a L-type brown dwarf.

\section{Conclusions} \label{Conclusion}

In this paper, we present and analyze a data set made from $R$=250\,000 PEPSI spectra taken continuously over 55 consecutive nights in 2022. These data were accompanied by $R$=55\,000 STELLA spectra taken contemporaneously over 522 nights in 2021--2022. The PEPSI data enabled for the first Doppler image of the surface of \lama. Our conclusions are as follows. 

Firstly, active-binary orbits are influenced by the RV modulation due to spots and must be reconsidered in case a more detailed comparison with dynamic evolutionary models is planned.   

Secondly, opposite to the RV models of solar-like stars \citep[e.g.,][]{Zhao2023}, the amplitude of the RV jitter on \lama\ is more than ten times larger than predicted and is accompanied by excess line broadening. We interpret the latter as anisotropic macro turbulence due to a spot-related moat flow and conclude that future Doppler-imaging studies should incorporate local velocity fields.  

Thirdly, \lama\ shows spatial coexistence of photospheric spots and the sources of chromospheric line emission. Seven spectral chromospheric tracers agree on a stable nature of its emission sources during our 522-d observing window. This also suggests that the underlying large-scale magnetic field morphology remains comparably stable as well. We conclude that, on the one hand, \lama\ may run a more chaotic dynamo compared to the Sun but, on the other hand, exhibits a fairly stable surface morphology. Whether this is a contradiction or being expected must be clarified in future dynamo simulations of active giant stars \citep[e.g.,][]{brun, inceoglu}. Besides, we did not detect any phase-coherent modulation of activity with the orbital period nor with the presence of a third body.

\begin{acknowledgements}
The authors thank the anonymous referee for the helpful comments that improved the quality of the article. This work is based partially on data obtained with the Stellar Activity-2 (STELLA-II) robotic telescope in Tenerife, an AIP facility jointly operated by AIP and IAC (https://stella.aip.de/). Also, partially on data from PEPSI acquired with the Large Binocular Telescope (LBT) and the Vatican Advanced Technology Telescope (VATT) (see https://pepsi.aip.de/). The LBT is an international collaboration among institutions in the United States, Italy and Germany. LBT Corporation partners are: The University of Arizona on behalf of the Arizona Board of Regents; Istituto Nazionale di Astrofisica, Italy; LBT Beteiligungsgesellschaft, Germany, representing the Max-Planck Society, The Leibniz Institute for Astrophysics Potsdam, and Heidelberg University; The Ohio State University, representing OSU, University of Notre Dame, University of Minnesota and University of Virginia.
In this work, we heavily used \texttt{python3} libraries; \texttt{astropy}\, \citep{astropy:2013, astropy:2018, astropy:2022},\, \texttt{numpy}\, \citep{numpyharris2020}\, and \texttt{scipy}\, \citep{SciPy2020-NMeth}.
The authors thank B. Seli from Konkoly Observatory for making available his Python code for spot segmentation \citep[see ][Appendix F]{Kovari2024}.
\"OA thanks S. P. J\"arvinen and T. A. Carroll, for sharing their experiences on the analyses of the observed data. 
ZsK acknowledges the financial support of the Hungarian National Research, Development and Innovation Office grant KKP-143986.
\end{acknowledgements}

\bibliographystyle{aa} 
\bibliography{ozgun} 

\newcommand{\noop}[1]{}
\begin{thebibliography}{77}
\expandafter\ifx\csname natexlab\endcsname\relax\def\natexlab#1{#1}\fi

\bibitem[{{Allende Prieto}(2004)}]{all04}
{Allende Prieto}, C. 2004, Astronomische Nachrichten, 325, 604

\bibitem[{{Allende Prieto} {et~al.}(2006){Allende Prieto}, {Beers}, {Wilhelm}, {Newberg}, {Rockosi}, {Yanny}, \& {Lee}}]{all}
{Allende Prieto}, C., {Beers}, T.~C., {Wilhelm}, R., {et~al.} 2006, \apj, 636, 804

\bibitem[{{Applegate}(1992)}]{Applegate1992}
{Applegate}, J.~H. 1992, \apj, 385, 621

\bibitem[{{Astropy Collaboration} {et~al.}(2022){Astropy Collaboration}, {Price-Whelan}, {Lim}, {Earl}, {Starkman}, {Bradley}, {Shupe}, {Patil}, {Corrales}, {Brasseur}, {N{"o}the}, {Donath}, {Tollerud}, {Morris}, {Ginsburg}, {Vaher}, {Weaver}, {Tocknell}, {Jamieson}, {van Kerkwijk}, {Robitaille}, {Merry}, {Bachetti}, {G{"u}nther}, {Aldcroft}, {Alvarado-Montes}, {Archibald}, {B{'o}di}, {Bapat}, {Barentsen}, {Baz{'a}n}, {Biswas}, {Boquien}, {Burke}, {Cara}, {Cara}, {Conroy}, {Conseil}, {Craig}, {Cross}, {Cruz}, {D'Eugenio}, {Dencheva}, {Devillepoix}, {Dietrich}, {Eigenbrot}, {Erben}, {Ferreira}, {Foreman-Mackey}, {Fox}, {Freij}, {Garg}, {Geda}, {Glattly}, {Gondhalekar}, {Gordon}, {Grant}, {Greenfield}, {Groener}, {Guest}, {Gurovich}, {Handberg}, {Hart}, {Hatfield-Dodds}, {Homeier}, {Hosseinzadeh}, {Jenness}, {Jones}, {Joseph}, {Kalmbach}, {Karamehmetoglu}, {Ka{l}uszy{'n}ski}, {Kelley}, {Kern}, {Kerzendorf}, {Koch}, {Kulumani}, {Lee}, {Ly}, {Ma}, {MacBride}, {Maljaars}, {Muna}, {Murphy}, {Norman}, {O'Steen},
  {Oman}, {Pacifici}, {Pascual}, {Pascual-Granado}, {Patil}, {Perren}, {Pickering}, {Rastogi}, {Roulston}, {Ryan}, {Rykoff}, {Sabater}, {Sakurikar}, {Salgado}, {Sanghi}, {Saunders}, {Savchenko}, {Schwardt}, {Seifert-Eckert}, {Shih}, {Jain}, {Shukla}, {Sick}, {Simpson}, {Singanamalla}, {Singer}, {Singhal}, {Sinha}, {Sip{H{o}}cz}, {Spitler}, {Stansby}, {Streicher}, {{{S}}umak}, {Swinbank}, {Taranu}, {Tewary}, {Tremblay}, {Val-Borro}, {Van Kooten}, {Vasovi{'c}}, {Verma}, {de Miranda Cardoso}, {Williams}, {Wilson}, {Winkel}, {Wood-Vasey}, {Xue}, {Yoachim}, {Zhang}, {Zonca}, \& {Astropy Project Contributors}}]{astropy:2022}
{Astropy Collaboration}, {Price-Whelan}, A.~M., {Lim}, P.~L., {et~al.} 2022, \apj, 935, 167

\bibitem[{{Astropy Collaboration} {et~al.}(2018){Astropy Collaboration}, {Price-Whelan}, {Sip{\H{o}}cz}, {G{\"u}nther}, {Lim}, {Crawford}, {Conseil}, {Shupe}, {Craig}, {Dencheva}, {Ginsburg}, {Vand erPlas}, {Bradley}, {P{\'e}rez-Su{\'a}rez}, {de Val-Borro}, {Aldcroft}, {Cruz}, {Robitaille}, {Tollerud}, {Ardelean}, {Babej}, {Bach}, {Bachetti}, {Bakanov}, {Bamford}, {Barentsen}, {Barmby}, {Baumbach}, {Berry}, {Biscani}, {Boquien}, {Bostroem}, {Bouma}, {Brammer}, {Bray}, {Breytenbach}, {Buddelmeijer}, {Burke}, {Calderone}, {Cano Rodr{\'\i}guez}, {Cara}, {Cardoso}, {Cheedella}, {Copin}, {Corrales}, {Crichton}, {D'Avella}, {Deil}, {Depagne}, {Dietrich}, {Donath}, {Droettboom}, {Earl}, {Erben}, {Fabbro}, {Ferreira}, {Finethy}, {Fox}, {Garrison}, {Gibbons}, {Goldstein}, {Gommers}, {Greco}, {Greenfield}, {Groener}, {Grollier}, {Hagen}, {Hirst}, {Homeier}, {Horton}, {Hosseinzadeh}, {Hu}, {Hunkeler}, {Ivezi{\'c}}, {Jain}, {Jenness}, {Kanarek}, {Kendrew}, {Kern}, {Kerzendorf}, {Khvalko}, {King}, {Kirkby}, {Kulkarni},
  {Kumar}, {Lee}, {Lenz}, {Littlefair}, {Ma}, {Macleod}, {Mastropietro}, {McCully}, {Montagnac}, {Morris}, {Mueller}, {Mumford}, {Muna}, {Murphy}, {Nelson}, {Nguyen}, {Ninan}, {N{\"o}the}, {Ogaz}, {Oh}, {Parejko}, {Parley}, {Pascual}, {Patil}, {Patil}, {Plunkett}, {Prochaska}, {Rastogi}, {Reddy Janga}, {Sabater}, {Sakurikar}, {Seifert}, {Sherbert}, {Sherwood-Taylor}, {Shih}, {Sick}, {Silbiger}, {Singanamalla}, {Singer}, {Sladen}, {Sooley}, {Sornarajah}, {Streicher}, {Teuben}, {Thomas}, {Tremblay}, {Turner}, {Terr{\'o}n}, {van Kerkwijk}, {de la Vega}, {Watkins}, {Weaver}, {Whitmore}, {Woillez}, {Zabalza}, \& {Astropy Contributors}}]{astropy:2018}
{Astropy Collaboration}, {Price-Whelan}, A.~M., {Sip{\H{o}}cz}, B.~M., {et~al.} 2018, \aj, 156, 123

\bibitem[{{Astropy Collaboration} {et~al.}(2013){Astropy Collaboration}, {Robitaille}, {Tollerud}, {Greenfield}, {Droettboom}, {Bray}, {Aldcroft}, {Davis}, {Ginsburg}, {Price-Whelan}, {Kerzendorf}, {Conley}, {Crighton}, {Barbary}, {Muna}, {Ferguson}, {Grollier}, {Parikh}, {Nair}, {Unther}, {Deil}, {Woillez}, {Conseil}, {Kramer}, {Turner}, {Singer}, {Fox}, {Weaver}, {Zabalza}, {Edwards}, {Azalee Bostroem}, {Burke}, {Casey}, {Crawford}, {Dencheva}, {Ely}, {Jenness}, {Labrie}, {Lim}, {Pierfederici}, {Pontzen}, {Ptak}, {Refsdal}, {Servillat}, \& {Streicher}}]{astropy:2013}
{Astropy Collaboration}, {Robitaille}, T.~P., {Tollerud}, E.~J., {et~al.} 2013, \aap, 558, A33

\bibitem[{{Baliunas} \& {Dupree}(1982)}]{Bal:Dup1982}
{Baliunas}, S.~L. \& {Dupree}, A.~K. 1982, \apj, 252, 668

\bibitem[{{Boisse} {et~al.}(2011){Boisse}, {Bouchy}, {H{\'e}brard}, {Bonfils}, {Santos}, \& {Vauclair}}]{Boisse2011}
{Boisse}, I., {Bouchy}, F., {H{\'e}brard}, G., {et~al.} 2011, \aap, 528, A4

\bibitem[{{Brun} \& {Palacios}(2009)}]{brun}
{Brun}, A.~S. \& {Palacios}, A. 2009, \apj, 702, 1078

\bibitem[{{Burns}(1906)}]{Burns1906}
{Burns}, K. 1906, \apj, 24, 345

\bibitem[{{Carlberg} {et~al.}(2012){Carlberg}, {Cunha}, {Smith}, \& {Majewski}}]{carl}
{Carlberg}, J.~K., {Cunha}, K., {Smith}, V.~V., \& {Majewski}, S.~R. 2012, \apj, 757, 109

\bibitem[{{Carroll} {et~al.}(2012){Carroll}, {Strassmeier}, {Rice}, \& {K{\"u}nstler}}]{carroll12}
{Carroll}, T.~A., {Strassmeier}, K.~G., {Rice}, J.~B., \& {K{\"u}nstler}, A. 2012, \aap, 548, A95

\bibitem[{{Castelli} \& {Kurucz}(2003)}]{atlas9}
{Castelli}, F. \& {Kurucz}, R.~L. 2003, in Modelling of Stellar Atmospheres, ed. N.~{Piskunov}, W.~W. {Weiss}, \& D.~F. {Gray}, Vol. 210, A20

\bibitem[{{Danby} \& {Burkardt}(1983)}]{dan:bur}
{Danby}, J.~M.~A. \& {Burkardt}, T.~M. 1983, Celestial Mechanics, 31, 95

\bibitem[{{Desort} {et~al.}(2007){Desort}, {Lagrange}, {Galland}, {Udry}, \& {Mayor}}]{Desort2007}
{Desort}, M., {Lagrange}, A.~M., {Galland}, F., {Udry}, S., \& {Mayor}, M. 2007, \aap, 473, 983

\bibitem[{{Donati} {et~al.}(1995){Donati}, {Henry}, \& {Hall}}]{Donati1995}
{Donati}, J.~F., {Henry}, G.~W., \& {Hall}, D.~S. 1995, \aap, 293, 107

\bibitem[{{Donati} {et~al.}(2016){Donati}, {Moutou}, {Malo}, {Baruteau}, {Yu}, {H{\'e}brard}, {Hussain}, {Alencar}, {M{\'e}nard}, {Bouvier}, {Petit}, {Takami}, {Doyon}, \& {Collier Cameron}}]{Donati2016}
{Donati}, J.~F., {Moutou}, C., {Malo}, L., {et~al.} 2016, \nat, 534, 662

\bibitem[{{Drake} {et~al.}(2011){Drake}, {Ball}, {Eldridge}, {Ness}, \& {Stancliffe}}]{Drake2011}
{Drake}, J.~J., {Ball}, B., {Eldridge}, J.~J., {Ness}, J.~U., \& {Stancliffe}, R.~J. 2011, \aj, 142, 144

\bibitem[{{Ducati}(2002)}]{Ducati2002}
{Ducati}, J.~R. 2002, {VizieR Online Data Catalog: Catalogue of Stellar Photometry in Johnson's 11-color system.}, CDS/ADC Collection of Electronic Catalogues, 2237, 0 (2002)

\bibitem[{{Gaia Collaboration} {et~al.}(2023){Gaia Collaboration}, {Vallenari}, {Brown}, {Prusti}, {de Bruijne}, {Arenou}, {Babusiaux}, {Biermann}, {Creevey}, {Ducourant}, {Evans}, {Eyer}, {Guerra}, {Hutton}, {Jordi}, {Klioner}, {Lammers}, {Lindegren}, {Luri}, {Mignard}, {Panem}, {Pourbaix}, {Randich}, {Sartoretti}, {Soubiran}, {Tanga}, {Walton}, {Bailer-Jones}, {Bastian}, {Drimmel}, {Jansen}, {Katz}, {Lattanzi}, {van Leeuwen}, {Bakker}, {Cacciari}, {Casta{\~n}eda}, {De Angeli}, {Fabricius}, {Fouesneau}, {Fr{\'e}mat}, {Galluccio}, {Guerrier}, {Heiter}, {Masana}, {Messineo}, {Mowlavi}, {Nicolas}, {Nienartowicz}, {Pailler}, {Panuzzo}, {Riclet}, {Roux}, {Seabroke}, {Sordo}, {Th{\'e}venin}, {Gracia-Abril}, {Portell}, {Teyssier}, {Altmann}, {Andrae}, {Audard}, {Bellas-Velidis}, {Benson}, {Berthier}, {Blomme}, {Burgess}, {Busonero}, {Busso}, {C{\'a}novas}, {Carry}, {Cellino}, {Cheek}, {Clementini}, {Damerdji}, {Davidson}, {de Teodoro}, {Nu{\~n}ez Campos}, {Delchambre}, {Dell'Oro}, {Esquej},
  {Fern{\'a}ndez-Hern{\'a}ndez}, {Fraile}, {Garabato}, {Garc{\'\i}a-Lario}, {Gosset}, {Haigron}, {Halbwachs}, {Hambly}, {Harrison}, {Hern{\'a}ndez}, {Hestroffer}, {Hodgkin}, {Holl}, {Jan{\ss}en}, {Jevardat de Fombelle}, {Jordan}, {Krone-Martins}, {Lanzafame}, {L{\"o}ffler}, {Marchal}, {Marrese}, {Moitinho}, {Muinonen}, {Osborne}, {Pancino}, {Pauwels}, {Recio-Blanco}, {Reyl{\'e}}, {Riello}, {Rimoldini}, {Roegiers}, {Rybizki}, {Sarro}, {Siopis}, {Smith}, {Sozzetti}, {Utrilla}, {van Leeuwen}, {Abbas}, {{\'A}brah{\'a}m}, {Abreu Aramburu}, {Aerts}, {Aguado}, {Ajaj}, {Aldea-Montero}, {Altavilla}, {{\'A}lvarez}, {Alves}, {Anders}, {Anderson}, {Anglada Varela}, {Antoja}, {Baines}, {Baker}, {Balaguer-N{\'u}{\~n}ez}, {Balbinot}, {Balog}, {Barache}, {Barbato}, {Barros}, {Barstow}, {Bartolom{\'e}}, {Bassilana}, {Bauchet}, {Becciani}, {Bellazzini}, {Berihuete}, {Bernet}, {Bertone}, {Bianchi}, {Binnenfeld}, {Blanco-Cuaresma}, {Blazere}, {Boch}, {Bombrun}, {Bossini}, {Bouquillon}, {Bragaglia}, {Bramante}, {Breedt},
  {Bressan}, {Brouillet}, {Brugaletta}, {Bucciarelli}, {Burlacu}, {Butkevich}, {Buzzi}, {Caffau}, {Cancelliere}, {Cantat-Gaudin}, {Carballo}, {Carlucci}, {Carnerero}, {Carrasco}, {Casamiquela}, {Castellani}, {Castro-Ginard}, {Chaoul}, {Charlot}, {Chemin}, {Chiaramida}, {Chiavassa}, {Chornay}, {Comoretto}, {Contursi}, {Cooper}, {Cornez}, {Cowell}, {Crifo}, {Cropper}, {Crosta}, {Crowley}, {Dafonte}, {Dapergolas}, {David}, {David}, {de Laverny}, {De Luise}, {De March}, {De Ridder}, {de Souza}, {de Torres}, {del Peloso}, {del Pozo}, {Delbo}, {Delgado}, {Delisle}, {Demouchy}, {Dharmawardena}, {Di Matteo}, {Diakite}, {Diener}, {Distefano}, {Dolding}, {Edvardsson}, {Enke}, {Fabre}, {Fabrizio}, {Faigler}, {Fedorets}, {Fernique}, {Fienga}, {Figueras}, {Fournier}, {Fouron}, {Fragkoudi}, {Gai}, {Garcia-Gutierrez}, {Garcia-Reinaldos}, {Garc{\'\i}a-Torres}, {Garofalo}, {Gavel}, {Gavras}, {Gerlach}, {Geyer}, {Giacobbe}, {Gilmore}, {Girona}, {Giuffrida}, {Gomel}, {Gomez}, {Gonz{\'a}lez-N{\'u}{\~n}ez},
  {Gonz{\'a}lez-Santamar{\'\i}a}, {Gonz{\'a}lez-Vidal}, {Granvik}, {Guillout}, {Guiraud}, {Guti{\'e}rrez-S{\'a}nchez}, {Guy}, {Hatzidimitriou}, {Hauser}, {Haywood}, {Helmer}, {Helmi}, {Sarmiento}, {Hidalgo}, {Hilger}, {H{\l}adczuk}, {Hobbs}, {Holland}, {Huckle}, {Jardine}, {Jasniewicz}, {Jean-Antoine Piccolo}, {Jim{\'e}nez-Arranz}, {Jorissen}, {Juaristi Campillo}, {Julbe}, {Karbevska}, {Kervella}, {Khanna}, {Kontizas}, {Kordopatis}, {Korn}, {K{\'o}sp{\'a}l}, {Kostrzewa-Rutkowska}, {Kruszy{\'n}ska}, {Kun}, {Laizeau}, {Lambert}, {Lanza}, {Lasne}, {Le Campion}, {Lebreton}, {Lebzelter}, {Leccia}, {Leclerc}, {Lecoeur-Taibi}, {Liao}, {Licata}, {Lindstr{\o}m}, {Lister}, {Livanou}, {Lobel}, {Lorca}, {Loup}, {Madrero Pardo}, {Magdaleno Romeo}, {Managau}, {Mann}, {Manteiga}, {Marchant}, {Marconi}, {Marcos}, {Marcos Santos}, {Mar{\'\i}n Pina}, {Marinoni}, {Marocco}, {Marshall}, {Martin Polo}, {Mart{\'\i}n-Fleitas}, {Marton}, {Mary}, {Masip}, {Massari}, {Mastrobuono-Battisti}, {Mazeh}, {McMillan}, {Messina}, {Michalik},
  {Millar}, {Mints}, {Molina}, {Molinaro}, {Moln{\'a}r}, {Monari}, {Mongui{\'o}}, {Montegriffo}, {Montero}, {Mor}, {Mora}, {Morbidelli}, {Morel}, {Morris}, {Muraveva}, {Murphy}, {Musella}, {Nagy}, {Noval}, {Oca{\~n}a}, {Ogden}, {Ordenovic}, {Osinde}, {Pagani}, {Pagano}, {Palaversa}, {Palicio}, {Pallas-Quintela}, {Panahi}, {Payne-Wardenaar}, {Pe{\~n}alosa Esteller}, {Penttil{\"a}}, {Pichon}, {Piersimoni}, {Pineau}, {Plachy}, {Plum}, {Poggio}, {Pr{\v{s}}a}, {Pulone}, {Racero}, {Ragaini}, {Rainer}, {Raiteri}, {Rambaux}, {Ramos}, {Ramos-Lerate}, {Re Fiorentin}, {Regibo}, {Richards}, {Rios Diaz}, {Ripepi}, {Riva}, {Rix}, {Rixon}, {Robichon}, {Robin}, {Robin}, {Roelens}, {Rogues}, {Rohrbasser}, {Romero-G{\'o}mez}, {Rowell}, {Royer}, {Ruz Mieres}, {Rybicki}, {Sadowski}, {S{\'a}ez N{\'u}{\~n}ez}, {Sagrist{\`a} Sell{\'e}s}, {Sahlmann}, {Salguero}, {Samaras}, {Sanchez Gimenez}, {Sanna}, {Santove{\~n}a}, {Sarasso}, {Schultheis}, {Sciacca}, {Segol}, {Segovia}, {S{\'e}gransan}, {Semeux}, {Shahaf}, {Siddiqui}, {Siebert},
  {Siltala}, {Silvelo}, {Slezak}, {Slezak}, {Smart}, {Snaith}, {Solano}, {Solitro}, {Souami}, {Souchay}, {Spagna}, {Spina}, {Spoto}, {Steele}, {Steidelm{\"u}ller}, {Stephenson}, {S{\"u}veges}, {Surdej}, {Szabados}, {Szegedi-Elek}, {Taris}, {Taylor}, {Teixeira}, {Tolomei}, {Tonello}, {Torra}, {Torra}, {Torralba Elipe}, {Trabucchi}, {Tsounis}, {Turon}, {Ulla}, {Unger}, {Vaillant}, {van Dillen}, {van Reeven}, {Vanel}, {Vecchiato}, {Viala}, {Vicente}, {Voutsinas}, {Weiler}, {Wevers}, {Wyrzykowski}, {Yoldas}, {Yvard}, {Zhao}, {Zorec}, {Zucker}, \& {Zwitter}}]{DR3}
{Gaia Collaboration}, {Vallenari}, A., {Brown}, A.~G.~A., {et~al.} 2023, \aap, 674, A1

\bibitem[{{Garraffo} {et~al.}(2015){Garraffo}, {Drake}, \& {Cohen}}]{Garraffo2015}
{Garraffo}, C., {Drake}, J.~J., \& {Cohen}, O. 2015, \apj, 813, 40

\bibitem[{{Garraffo} {et~al.}(2018){Garraffo}, {Drake}, {Dotter}, {Choi}, {Burke}, {Moschou}, {Alvarado-G{\'o}mez}, {Kashyap}, \& {Cohen}}]{Garraffo2018}
{Garraffo}, C., {Drake}, J.~J., {Dotter}, A., {et~al.} 2018, \apj, 862, 90

\bibitem[{{Gustafsson} {et~al.}(2008){Gustafsson}, {Edvardsson}, {Eriksson}, {J{\o}rgensen}, {Nordlund}, \& {Plez}}]{Marcs2008}
{Gustafsson}, B., {Edvardsson}, B., {Eriksson}, K., {et~al.} 2008, \aap, 486, 951

\bibitem[{{Hall}(1972)}]{Hall1972}
{Hall}, D.~S. 1972, \pasp, 84, 323

\bibitem[{{Hall}(1976)}]{Hall1976}
{Hall}, D.~S. 1976, in Astrophysics and Space Science Library, Vol.~60, IAU Colloq. 29: Multiple Periodic Variable Stars, ed. W.~S. {Fitch}, 287

\bibitem[{{Hall}(1996)}]{Hall1996}
{Hall}, J.~C. 1996, \pasp, 108, 313

\bibitem[{Harris {et~al.}(2020)Harris, Millman, van~der Walt, Gommers, Virtanen, Cournapeau, Wieser, Taylor, Berg, Smith, Kern, Picus, Hoyer, van Kerkwijk, Brett, Haldane, del R{\'{i}}o, Wiebe, Peterson, G{\'{e}}rard-Marchant, Sheppard, Reddy, Weckesser, Abbasi, Gohlke, \& Oliphant}]{numpyharris2020}
Harris, C.~R., Millman, K.~J., van~der Walt, S.~J., {et~al.} 2020, Nature, 585, 357

\bibitem[{{Hatzes}(2002)}]{Hatzes2002}
{Hatzes}, A.~P. 2002, Astronomische Nachrichten, 323, 392

\bibitem[{{Henry} {et~al.}(1995){Henry}, {Eaton}, {Hamer}, \& {Hall}}]{Henry1995}
{Henry}, G.~W., {Eaton}, J.~A., {Hamer}, J., \& {Hall}, D.~S. 1995, \apjs, 97, 513

\bibitem[{{Husser} {et~al.}(2013){Husser}, {Wende-von Berg}, {Dreizler}, {Homeier}, {Reiners}, {Barman}, \& {Hauschildt}}]{Husser2013}
{Husser}, T.~O., {Wende-von Berg}, S., {Dreizler}, S., {et~al.} 2013, \aap, 553, A6

\bibitem[{{Hut}(1981)}]{Hut1981}
{Hut}, P. 1981, \aap, 99, 126

\bibitem[{{Ilyin}(2000)}]{ilyin4A}
{Ilyin}, I. 2000, {PhD Thesis}

\bibitem[{{Ilyin}(2012)}]{ilyin2012}
{Ilyin}, I. 2012, Astronomische Nachrichten, 333, 213

\bibitem[{{Inceoglu} {et~al.}(2019){Inceoglu}, {Simoniello}, {Arlt}, \& {Rempel}}]{inceoglu}
{Inceoglu}, F., {Simoniello}, R., {Arlt}, R., \& {Rempel}, M. 2019, \aap, 625, A117

\bibitem[{{J{\"a}rvinen} {et~al.}(2018){J{\"a}rvinen}, {Strassmeier}, {Carroll}, {Ilyin}, \& {Weber}}]{ekdra}
{J{\"a}rvinen}, S.~P., {Strassmeier}, K.~G., {Carroll}, T.~A., {Ilyin}, I., \& {Weber}, M. 2018, \aap, 620, A162

\bibitem[{{Jovanovic} {et~al.}(2013){Jovanovic}, {Weber}, \& {Allende Prieto}}]{parses}
{Jovanovic}, M., {Weber}, M., \& {Allende Prieto}, C. 2013, Publications de l'Observatoire Astronomique de Beograd, 92, 169

\bibitem[{{K{\H{o}}v{\'a}ri} {et~al.}(2024){K{\H{o}}v{\'a}ri}, {Strassmeier}, {Kriskovics}, {Ol{\'a}h}, {Borkovits}, {Radv{\'a}nyi}, {Granzer}, {Seli}, {Vida}, \& {Weber}}]{Kovari2024}
{K{\H{o}}v{\'a}ri}, Z., {Strassmeier}, K.~G., {Kriskovics}, L., {et~al.} 2024, \aap, 684, A94

\bibitem[{{Kochukhov} {et~al.}(2010){Kochukhov}, {Makaganiuk}, \& {Piskunov}}]{Kochukhov2010}
{Kochukhov}, O., {Makaganiuk}, V., \& {Piskunov}, N. 2010, \aap, 524, A5

\bibitem[{{Kurucz}(1993)}]{atlas9grid}
{Kurucz}, R. 1993, ATLAS9 Stellar Atmosphere Programs and 2 km/s grid. Kurucz CD-ROM No. 13. Cambridge, 13

\bibitem[{{Landis} {et~al.}(1978){Landis}, {Lovell}, {Hall}, {Henry}, \& {Renner}}]{Landis1978}
{Landis}, H.~J., {Lovell}, L.~P., {Hall}, D.~S., {Henry}, G.~W., \& {Renner}, T.~R. 1978, \aj, 83, 176

\bibitem[{{Lomb}(1976)}]{Lomb1976}
{Lomb}, N.~R. 1976, \apss, 39, 447

\bibitem[{{MacLeod} {et~al.}(2025){MacLeod}, {Blunt}, {De Rosa}, {Dupree}, {Granzer}, {Harper}, {Huang}, {Leiner}, {Loeb}, {Nielsen}, {Strassmeier}, {Wang}, \& {Weber}}]{MacLeod2025}
{MacLeod}, M., {Blunt}, S., {De Rosa}, R.~J., {et~al.} 2025, \apj, 978, 50

\bibitem[{{Martinez} {et~al.}(2021){Martinez}, {Baron}, {Monnier}, {Roettenbacher}, \& {Parks}}]{Martinez2021}
{Martinez}, A.~O., {Baron}, F.~R., {Monnier}, J.~D., {Roettenbacher}, R.~M., \& {Parks}, J.~R. 2021, \apj, 916, 60

\bibitem[{{Massarotti} {et~al.}(2008){Massarotti}, {Latham}, {Stefanik}, \& {Fogel}}]{Massarotti2008}
{Massarotti}, A., {Latham}, D.~W., {Stefanik}, R.~P., \& {Fogel}, J. 2008, \aj, 135, 209

\bibitem[{{Menuier}(2023)}]{Meunier2023}
{Menuier}, N. 2023, in Star-Planet Interactions, ed. L.~{Bigot}, J.~{Bouvier}, Y.~{Lebreton}, A.~{Chiavassa}, \& A.~{L{\`e}bre}, 22

\bibitem[{{Mirtorabi} {et~al.}(2003){Mirtorabi}, {Wasatonic}, \& {Guinan}}]{Mirtorabi2003}
{Mirtorabi}, M.~T., {Wasatonic}, R., \& {Guinan}, E.~F. 2003, \aj, 125, 3265

\bibitem[{{{\'O} Fionnag{\'a}in} {et~al.}(2021){{\'O} Fionnag{\'a}in}, {Vidotto}, {Petit}, {Neiner}, {Manchester}, {Folsom}, \& {Hallinan}}]{Fionnagain2021}
{{\'O} Fionnag{\'a}in}, D., {Vidotto}, A.~A., {Petit}, P., {et~al.} 2021, \mnras, 500, 3438

\bibitem[{{Parks} {et~al.}(2021){Parks}, {White}, {Baron}, {Monnier}, {Kloppenborg}, {Henry}, {Schaefer}, {Che}, {Pedretti}, {Thureau}, {Zhao}, {ten Brummelaar}, {McAlister}, {Ridgway}, {Turner}, {Sturmann}, \& {Sturmann}}]{Parks2021}
{Parks}, J.~R., {White}, R.~J., {Baron}, F., {et~al.} 2021, \apj, 913, 54

\bibitem[{{Piskunov} \& {Wehlau}(1990)}]{pis:weh}
{Piskunov}, N.~E. \& {Wehlau}, W.~H. 1990, \aap, 233, 497

\bibitem[{{Plez}(2012)}]{turbo}
{Plez}, B. 2012, {Turbospectrum: Code for spectral synthesis}, Astrophysics Source Code Library, record ascl:1205.004

\bibitem[{{Popper}(1980)}]{Popper1980}
{Popper}, D.~M. 1980, \araa, 18, 115

\bibitem[{{Rajpaul} {et~al.}(2015){Rajpaul}, {Aigrain}, {Osborne}, {Reece}, \& {Roberts}}]{Rajpaul2015}
{Rajpaul}, V., {Aigrain}, S., {Osborne}, M.~A., {Reece}, S., \& {Roberts}, S. 2015, \mnras, 452, 2269

\bibitem[{{Ryabchikova} {et~al.}(2015){Ryabchikova}, {Piskunov}, {Kurucz}, {Stempels}, {Heiter}, {Pakhomov}, \& {Barklem}}]{vald3}
{Ryabchikova}, T., {Piskunov}, N., {Kurucz}, R.~L., {et~al.} 2015, \physscr, 90, 054005

\bibitem[{{Saar} \& {Donahue}(1997)}]{Saar&Donahue1997}
{Saar}, S.~H. \& {Donahue}, R.~A. 1997, \apj, 485, 319

\bibitem[{{Savanov} \& {Berdyugina}(1994)}]{Savanov&Berdyugina1994}
{Savanov}, I.~S. \& {Berdyugina}, S.~V. 1994, Astronomy Letters, 20, 227

\bibitem[{{Scargle}(1982)}]{Scargle1982}
{Scargle}, J.~D. 1982, \apj, 263, 835

\bibitem[{{Solanki}(2003)}]{Solanki2003}
{Solanki}, S.~K. 2003, \aapr, 11, 153

\bibitem[{{Solanki} \& {Stenflo}(1984)}]{Sol&Ste1984}
{Solanki}, S.~K. \& {Stenflo}, J.~O. 1984, \aap, 140, 185

\bibitem[{{Stellingwerf}(1978)}]{pdm}
{Stellingwerf}, R.~F. 1978, \apj, 224, 953

\bibitem[{{Strassmeier}(2009)}]{Strassmeier2009}
{Strassmeier}, K.~G. 2009, \aapr, 17, 251

\bibitem[{{Strassmeier} {et~al.}(2019){Strassmeier}, {Carroll}, \& {Ilyin}}]{Strassmeier2019}
{Strassmeier}, K.~G., {Carroll}, T.~A., \& {Ilyin}, I.~V. 2019, \aap, 625, A27

\bibitem[{{Strassmeier} {et~al.}(2004){Strassmeier}, {Granzer}, {Weber}, {Woche}, {Andersen}, {Bartus}, {Bauer}, {Dionies}, {Popow}, {Fechner}, {Hildebrandt}, {Washuettl}, {Ritter}, {Schwope}, {Staude}, {Paschke}, {Stolz}, {Serre-Ricart}, {de la Rosa}, \& {Arnay}}]{stella2004}
{Strassmeier}, K.~G., {Granzer}, T., {Weber}, M., {et~al.} 2004, Astronomische Nachrichten, 325, 527

\bibitem[{{Strassmeier} {et~al.}(1989){Strassmeier}, {Hall}, {Boyd}, \& {Genet}}]{Strassmeier1989}
{Strassmeier}, K.~G., {Hall}, D.~S., {Boyd}, L.~J., \& {Genet}, R.~M. 1989, \apjs, 69, 141

\bibitem[{{Strassmeier} {et~al.}(1993){Strassmeier}, {Hall}, {Fekel}, \& {Scheck}}]{Strassmeier1993}
{Strassmeier}, K.~G., {Hall}, D.~S., {Fekel}, F.~C., \& {Scheck}, M. 1993, \aaps, 100, 173

\bibitem[{{Strassmeier} {et~al.}(2015){Strassmeier}, {Ilyin}, {J{\"a}rvinen}, {Weber}, {Woche}, {Barnes}, {Bauer}, {Beckert}, {Bittner}, {Bredthauer}, {Carroll}, {Denker}, {Dionies}, {DiVarano}, {D{\"o}scher}, {Fechner}, {Feuerstein}, {Granzer}, {Hahn}, {Harnisch}, {Hofmann}, {Lesser}, {Paschke}, {Pankratow}, {Plank}, {Pl{\"u}schke}, {Popow}, \& {Sablowski}}]{Strassmeier2015}
{Strassmeier}, K.~G., {Ilyin}, I., {J{\"a}rvinen}, A., {et~al.} 2015, Astronomische Nachrichten, 336, 324

\bibitem[{{Strassmeier} {et~al.}(2018){Strassmeier}, {Ilyin}, \& {Steffen}}]{Str2018}
{Strassmeier}, K.~G., {Ilyin}, I., \& {Steffen}, M. 2018, \aap, 612, A44

\bibitem[{{Strassmeier} {et~al.}(2024){Strassmeier}, {Ilyin}, {Woche}, {Dionies}, {Weber}, {J{\"a}rvinen}, {Denker}, {Dineva}, {Verma}, {Granzer}, {Bittner}, {Bauer}, {Paschke}, \& {{\~A}-nel}}]{sdipol}
{Strassmeier}, K.~G., {Ilyin}, I., {Woche}, M., {et~al.} 2024, Astronomische Nachrichten, 345, e20240033

\bibitem[{{Strassmeier} \& {Steffen}(2022)}]{Strassmeier2022}
{Strassmeier}, K.~G. \& {Steffen}, M. 2022, Astronomische Nachrichten, 343, e20220036

\bibitem[{{Strassmeier} {et~al.}(2023){Strassmeier}, {Weber}, {Gruner}, {Ilyin}, {Steffen}, {Baratella}, {J{\"a}rvinen}, {Granzer}, {Barnes}, {Carroll}, {Mallonn}, {Sablowski}, {Gabor}, {Brown}, {Corbally}, \& {Franz}}]{vpnep}
{Strassmeier}, K.~G., {Weber}, M., {Gruner}, D., {et~al.} 2023, \aap, 671, A7

\bibitem[{{Tautvai{\v{s}}ien{\.{e}}} {et~al.}(2010){Tautvai{\v{s}}ien{\.{e}}}, {Barisevi{\v{c}}ius}, {Berdyugina}, {Chorniy}, \& {Ilyin}}]{Tautvaisiene2010}
{Tautvai{\v{s}}ien{\.{e}}}, G., {Barisevi{\v{c}}ius}, G., {Berdyugina}, S., {Chorniy}, Y., \& {Ilyin}, I. 2010, Baltic Astronomy, 19, 95

\bibitem[{{Toner} \& {Gray}(1988)}]{Toner&Gray1988}
{Toner}, C.~G. \& {Gray}, D.~F. 1988, \apj, 334, 1008

\bibitem[{{van Leeuwen}(2007)}]{hip}
{van Leeuwen}, F. 2007, \aap, 474, 653

\bibitem[{Virtanen {et~al.}(2020)Virtanen, Gommers, Oliphant, Haberland, Reddy, Cournapeau, Burovski, Peterson, Weckesser, Bright, {van der Walt}, Brett, Wilson, Millman, Mayorov, Nelson, Jones, Kern, Larson, Carey, Polat, Feng, Moore, {VanderPlas}, Laxalde, Perktold, Cimrman, Henriksen, Quintero, Harris, Archibald, Ribeiro, Pedregosa, {van Mulbregt}, \& {SciPy 1.0 Contributors}}]{SciPy2020-NMeth}
Virtanen, P., Gommers, R., Oliphant, T.~E., {et~al.} 2020, Nature Methods, 17, 261

\bibitem[{{Vogt} \& {Penrod}(1983)}]{Vogt1983}
{Vogt}, S.~S. \& {Penrod}, G.~D. 1983, \pasp, 95, 565

\bibitem[{{Walker}(1944)}]{Walker1944}
{Walker}, E.~C. 1944, \jrasc, 38, 249

\bibitem[{{Weber} {et~al.}(2016){Weber}, {Granzer}, \& {Strassmeier}}]{weber16}
{Weber}, M., {Granzer}, T., \& {Strassmeier}, K.~G. 2016, in Society of Photo-Optical Instrumentation Engineers (SPIE) Conference Series, Vol. 9910, Observatory Operations: Strategies, Processes, and Systems VI, ed. A.~B. {Peck}, R.~L. {Seaman}, \& C.~R. {Benn}, 99100N

\bibitem[{{Zhao} \& {Dumusque}(2023)}]{Zhao2023}
{Zhao}, Y. \& {Dumusque}, X. 2023, \aap, 671, A11

\end{thebibliography}

\clearpage

\begin{appendix}
\onecolumn

\section{Extra tables}

\begin{table*}[h!]
\caption{Observing log for STELLA-SES data.}
\begin{tabular}{lcccccc}
\hline\hline\noalign{\smallskip}
Date (UT)  &BJD &Orbital Phase  & Velocity [\kms] & O-C [\ms] & Rotational Phase  \\
\noalign{\smallskip}\hline\noalign{\smallskip}
2021-08-09   &2459436.4728   &0.3345    &6.021   &166.40    &0.2235    \\
2021-08-11   &2459437.5456   &0.3868    &4.100   &93.81     &0.2432    \\
2021-08-12   &2459438.5085   &0.4337    &2.704   &71.24     &0.2609    \\
\dots \\
\dots \\
2023-01-09   &2459954.3831   &0.5734    &0.7671   &-6.3801     &0.7417    \\
2023-01-13   &2459958.3848   &0.7684    &4.7661   &47.0927     &0.8153    \\
2023-01-15   &2459960.4089   &0.8670    &8.9863   &135.0966    &0.8525    \\
\noalign{\smallskip}\hline
\end{tabular}
\tablefoot{Complete table is available at the CDS.}
\label{Table_STELLA-SES}
\end{table*}

\begin{table*}[h!]
\caption{Observing log for PEPSI-VATT data.}
\begin{tabular}{lcccccc}
\hline\hline\noalign{\smallskip}
Date (UT) & BJD &Orbital Phase & Velocity [\kms] & O-C [\ms] & Rotational Phase  \\
\noalign{\smallskip}\hline\noalign{\smallskip}
2022-05-10    &2459709.9697   &0.6610   &1.4508    &-192.41  &0.2499  \\
2022-05-10    &2459709.9832   &0.6626   &1.4597    &-196.13  &0.2501  \\
2022-05-11    &2459710.9395   &0.7099   &2.5146    &-253.60  &0.2677  \\
\dots \\                                                               
\dots \\
2022-06-30    &2459760.9569   &0.1473   &12.8141   &116.34   &0.1866  \\
2022-07-01    &2459761.9130   &0.1939   &11.3057   &20.42    &0.2045  \\
2022-07-01    &2459761.9439   &0.1954   &11.2133   &-20.20   &0.2050  \\
\noalign{\smallskip}\hline
\end{tabular}
\tablefoot{Complete table is available at the CDS.}
\label{Table_PEPSI-VATT}
\end{table*}

\begin{table}[h!]
\caption{Period analysis of \lama\ for different activity indicators.}\label{Period_Analysis}
\begin{tabular}{c r r r r r r r r c}
\hline\hline\noalign{\smallskip}
Activity              &$P_{\rm LS}$          &$P_{\rm PDM}$ \\
Indicator             &(days)            &(days)   \\
\noalign{\smallskip}\hline\noalign{\smallskip}
Ca\,{\sc ii} IRT-1    & 54.3 $\pm$ 0.1    &  54.4 $\pm$ 0.3   \\
Ca\,{\sc ii} IRT-2    & 54.3 $\pm$ 0.1    &  54.5 $\pm$ 0.2   \\
Ca\,{\sc ii} IRT-3    & 54.2 $\pm$ 0.1    &  54.4 $\pm$ 0.2   \\
Ca H                  & 54.2 $\pm$ 0.1    &  54.3 $\pm$ 0.3   \\
Ca K                  & 54.2 $\pm$ 0.1    &  54.3 $\pm$ 0.3   \\
\Halpha               & 53.8 $\pm$ 0.4    &  54.2 $\pm$ 0.5   \\
\Hbeta                & 52.7 $\pm$ 1.1    &  54.2 $\pm$ 1.4   \\
$\Delta$T             & 54.3 $\pm$ 0.1    &  54.3 $\pm$ 0.2   \\
RV residuals          & 54.4 $\pm$ 0.3    &  54.4 $\pm$ 0.4   \\
\noalign{\smallskip}\hline
\end{tabular}
\tablefoot{Rotational period is determined by using different activity indicators with two methods; Lomb-Scargle (LS) periodogram and phase dispersion minimization (PDM). They all agree $\sim$ 54 days. While Ca\,{\sc ii} IRT and \cahk\, lines demonstrate the least deviations from the period we determined by using the RV residuals, \Halpha\, and \Hbeta\, lines show higher deviations with larger error bars.}
\end{table}

\clearpage

\newpage
\twocolumn
\section{Extra figures}

\begin{figure}[h!]
    \includegraphics[width= 0.5\textwidth]{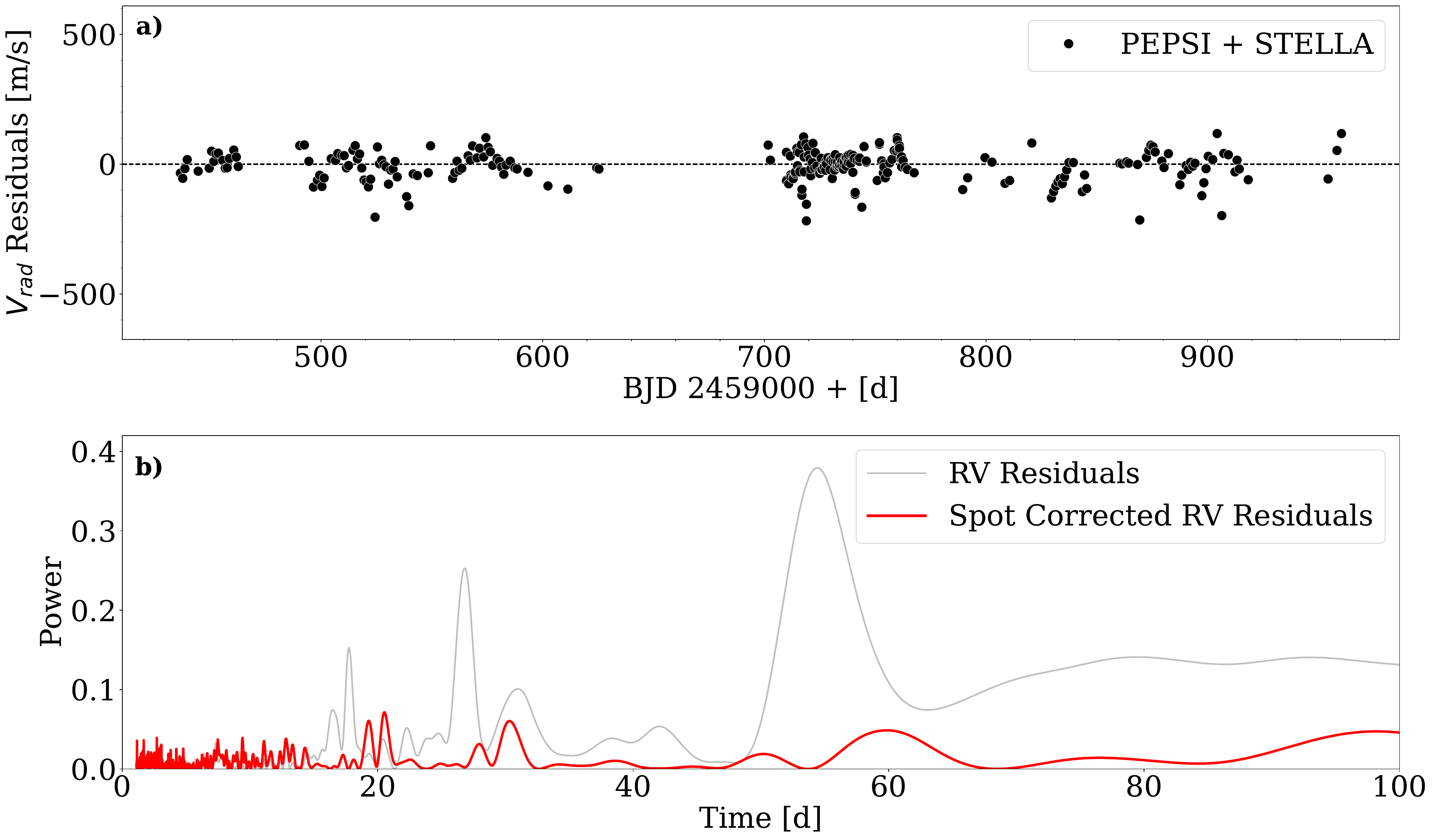}
    \caption{Panel \emph{a}: Orbital RV residuals after the spot correction. Panel \emph{b}: Lomb-Scargle periodogram of the residuals in panel \emph{a} (red line) compared with the periodogram before spot correction (gray line). The rotation period of the primary of 54.4\,d and its aliases are clearly seen from the residuals but are absent after the correction.}
    \label{RV_Res2}
\end{figure}


\begin{figure}[ht!]
    \includegraphics[width= 0.5\textwidth]{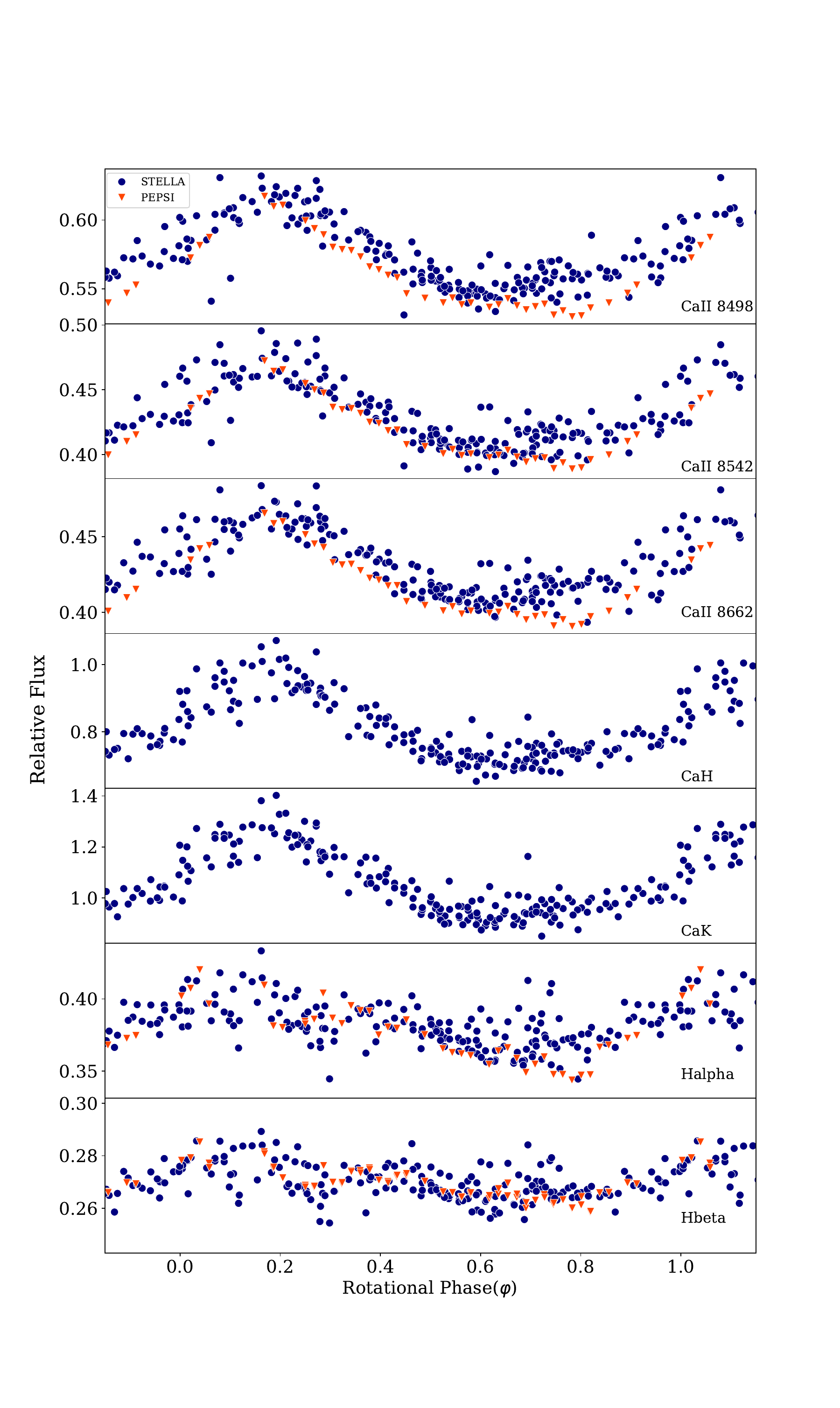}
    \caption{Phase folded chromospheric activity tracers for the time period January 2021 to July 2022. From top to bottom: \cairt, \cahk, \Halpha, and \Hbeta. Only relative fluxes are plotted.}
    \label{Tracers}
\end{figure}

\begin{figure*}[h!]
    \centering
    \includegraphics[width= 15cm]{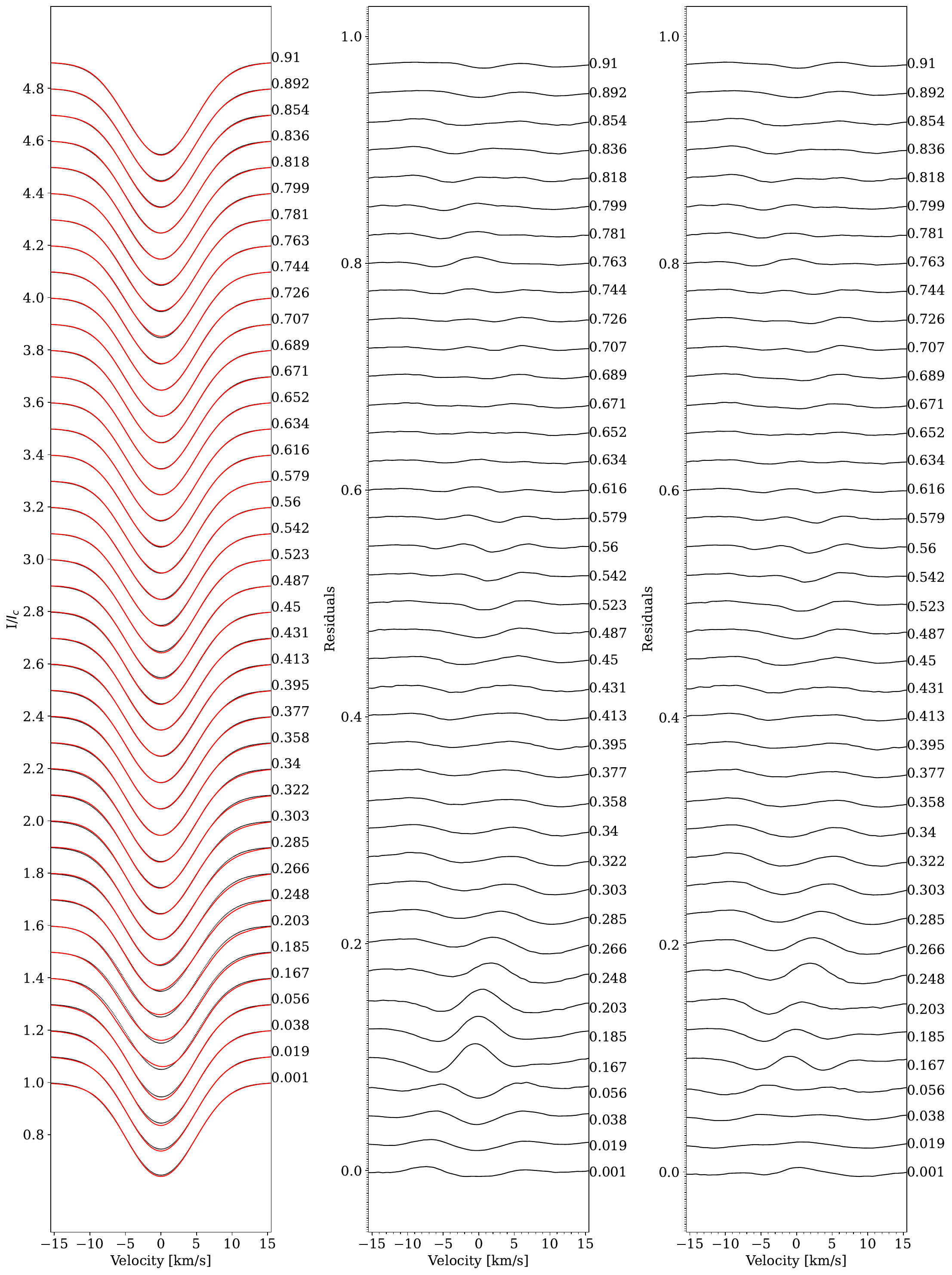}
    \caption{Singular value decomposition line profile fits from $i$MAP. \textit{Left panel:} Observed line profiles (red line) and synthetic profiles (black dotted lines). \textit{Middle panel:} Residuals from the $i$MAP fits with a single fixed macro turbulence. \textit{Right panel:} Residuals when a macro turbulence change is applied to the first seven phases (see text). Numbers on the right side of each panel indicate the rotational phase. }
    \label{SVDs}
\end{figure*}

\end{appendix}

\end{document}